\documentclass[a4paper,11pt]{article}
\pdfoutput=1
\usepackage[english]{babel}

\usepackage[a4paper,tmargin=3truecm,bmargin=3truecm,rmargin=3truecm,
lmargin=3truecm,twoside,verbose=true]{geometry}

\usepackage{cancel,graphicx}
\usepackage{hyperref}

\usepackage{amsmath,amssymb}

\numberwithin{equation}{section}
\allowdisplaybreaks[1]

\usepackage{cite}

\usepackage[amsmath, hyperref, thmmarks]{ntheorem}

\usepackage{cancel,graphicx}
\usepackage{hyperref}
\numberwithin{equation}{section}

\allowdisplaybreaks[1]

\newcommand{\be}{\begin{equation}}
\newcommand{\ee}{\end{equation}}

\newcommand\bbone{{ \mathbb{I}}}

\DeclareMathOperator{\tr}{Tr}

\newcommand{\deq}{: \hspace{-0.5pt} =}
\newcommand{\dreq}{= \hspace{-0.5pt} :}

\theoremsymbol{}
\theorembodyfont{\slshape}
\theoremheaderfont{\normalfont\bfseries}
\theoremseparator{}

\theorembodyfont{\upshape}
\theoremsymbol{\ensuremath{\blacklozenge}}

\theoremstyle{nonumberplain}
\theoremheaderfont{\scshape}
\theorembodyfont{\normalfont}
\theoremsymbol{\ensuremath{\blacksquare}}

\qedsymbol{\ensuremath{_\blacksquare}}
\theoremclass{LaTeX}
\renewenvironment{thebibliography}[1]
         {\section*{References}\frenchspacing\small
          \begin{list}{[\arabic{enumi}]}
         {\usecounter{enumi}\parsep=2pt\topsep 0pt
         \settowidth{\labelwidth}{[#1]}
         \leftmargin=\labelwidth\advance\leftmargin\labelsep
         \rightmargin=0pt\itemsep=1pt\sloppy}}{\end{list}}
\title{Involutive representations of coordinate algebras \\
and quantum spaces}

\author{Tajron Juri\'c$^{a,b}$, Timoth\'e Poulain$^c$, Jean-Christophe Wallet$^c$}

\begin{document}
\date{}
\maketitle
\begin{center}
\textit{$^a$Ru\dj er Bo\v{s}kovi\'c Institute, Theoretical Physics Division\\
Bijeni\v{c}ka c.54, HR-10002 Zagreb, Croatia\\
$^b$Instituto de Fisica, Universidade de Brasilia,\\
Caixa Postal 04455, 70919-970, Brasilia, DF, Brazil}\\
\href{mailto:tjuric@irb.hr}{\texttt{tjuric@irb.hr}}\\[1ex]%
\textit{$^c$Laboratoire de Physique Th\'eorique, B\^at.\ 210\\
CNRS and Universit\'e Paris-Sud 11,  91405 Orsay Cedex, France}\\
 \href{mailto:timothe.poulain@th.u-psud.fr}{\texttt{timothe.poulain@th.u-psud.fr}}, \href{mailto:jean-christophe.wallet@th.u-psud.fr}{\texttt{jean-christophe.wallet@th.u-psud.fr}}\\[1ex]%
\end{center}


\maketitle


\begin{abstract} 
We show that $\mathfrak{su}(2)$ Lie algebras of coordinate operators related to quantum spaces with $\mathfrak{su}(2)$ noncommutativity can be conveniently represented by $SO(3)$-covariant poly-differential involutive representations. We show that the quantized plane waves obtained from the quantization map action on the usual exponential functions are determined by polar decomposition of operators combined with constraint stemming from the Wigner theorem for $SU(2)$. Selecting a subfamily of $^*$-representations, we show that the resulting star-product is equivalent to the Kontsevich product for the Poisson manifold dual to the finite dimensional Lie algebra $\mathfrak{su}(2)$. We discuss the results, indicating a way to extend the construction to any semi-simple non simply connected Lie group and present noncommutative scalar field theories which are free from perturbative UV/IR mixing.
\end{abstract}

\pagebreak

\section{Introduction}

Noncommutative Geometry \cite{Connes} provides a way to resolve the physical objections 
to the co-existence of continuous space-time and commuting coordinates at the Planck scale \cite{Doplich1}. Noncommutativity of space-time might have significant physical impact, e.g. gravitational and cosmological effects as investigated in \cite{cosmogol1,cosmogol2,cosmogol3}. Once the noncommutative nature of space-time is assumed which is generally reflected into the noncommutativity of an algebra of coordinates, noncommutative field theories (NCFT) or noncommutative gauge theories naturally arise. For reviews on early studies, see e.g \cite{dnsw-rev} and references therein. \\

One way to investigate quantum properties of a NCFT is to represent it as a matrix model. This has been done in \cite{Grosse:2003aj-pc,matrix1,matrix2,vitwal} for the Moyal spaces and for deformations of $\mathbb{R}^3$ which define a class of noncommutative spaces, or quantum spaces, with $\mathfrak{su}(2)$ noncommutativity. These latter spaces have been introduced a long time ago in \cite{lagraa} (see also \cite{selene,rosa-vit,vit-kust}). For an alternative description in terms of convolution algebra of $SU(2)$, see \cite{wal-16}. These spaces are of our main concern in the present paper. In the matrix model approach of NCFT, one regards the noncommutative coordinate algebras as (abstract) involutive algebras of (self-adjoint) operators so that standard tools and properties of functional analysis can be exploited to study quantum properties of the NCFT. Recall that an involutive algebra $\mathbb{A}$, also known as $^*$-algebra, is an algebra endowed with an involution, i.e a map $\mathcal{I}:\mathbb{A}\to\mathbb{A}$ such that $\mathcal{I}^2(a)=a$, for any $a\in\mathbb{A}$. Albeit powerful, it may happen that the matrix model formulation of a NCFT becomes unexploitable due to severe technical difficulties.\\

Another alternative (widely used) framework to study properties of NCFT is to take advantage of star-products and deformation theory approach, either from the standard viewpoint of formal deformations extending the quantization approach of classical phase spaces \cite{deform}, or taking advantage of underlying Hopf algebra structures and related twists \cite{leningrad}. We adopt the former viewpoint in this paper. Informally, this framework amounts (among other tasks) to represent the abstract involutive algebras of operators stemming from the above mentioned coordinate algebras on well chosen {\it{involutive}} algebras of functions equipped with a deformed product, i.e star-product, which can be achieved through the introduction of some suitable (invertible) quantization map. One well-known example heavily used in earlier studies of quantum properties of NCFT on Moyal spaces \cite{Wallet:2007c,Blaschke:2009c} is the Weyl quantization map, linked to the Wigner-Weyl transform, giving rise to the Moyal product. Other star-products related to $\kappa$-Minkowski spaces as well as deformations of $\mathbb{R}^3$ with $\mathfrak{su}(2)$ noncommutativity have also appeared and used to construct and study NCFT on these spaces \cite{cosmogol4,kappa-star1,kappa-star,jpw-16,KV-15,polydiff-alg,VK,cosmogol5}. The use of star-product formulation of NCFT is often convenient for fast construction of reasonable functional actions but may lead to technical difficulties whenever the star-product is represented by a complicated formula and/or is not closed for a trace functional. \\

In this paper, we will focus on quantum spaces with $\mathfrak{su}(2)$ noncommutativity. Since these spaces support a canonical Poisson structure stemming from the underlying $\mathfrak{su}(2)$ Lie algebra structure, the general framework for the deformations of Poisson manifolds \cite{deform} could be applied to the present situation. However, this would lead to star-products expressed as infinite expansions which cannot be used to study quantum properties of NCFT. Explicit (closed) expressions for these star-products are needed. A convenient way to reach this goal is to represent the abstract $^*$-algebra of coordinate operators as $^*$-algebra of differential operators. Therefore, one natural requirement for these differential representations is to be compatible with the involution. Recall that for such an involutive representation, or $^*$-representation, says $\pi:\mathbb{A}\to\mathbb{B}$ a morphism of algebras ($\mathbb{A}$ and $\mathbb{B}$ are $^*$-algebras with respective involutions $*$ and $\dag$), one has $\pi(a^*)=(\pi(a))^\dag$. This is what we demand for the differential representations we look for. Differential representations have been used in various contexts in the Physics literature, see e.g. \cite{cosmogol4,kappa-star1,kappa-star,jpw-16,KV-15,polydiff-alg,VK,cosmogol5} and references therein, some being $^*$-representations, some being not. The use of a non $^*$-representation may lead to troublesome features. For instance, self-adjoint coordinate operators may not be represented as self-adjoint differential operators, unitary operators built from the exponential map may not be represented as unitary differential operators. Besides, one may obtain a star-product for which, in obvious notations, $(f\star g)^\dag\ne g^\dag \star f^\dag$ (for any $f$ and $g$ in some suitable space of functions) what may be problematic to define natural reality conditions on functional actions.\\

Before entering the discussion, it is instructive to compare briefly our present approach to another approach based on a suitable (noncommutative) adaptation of the Hopf map defining the well known Hopf fibration of the 3-sphere $\mathbb{S}_3$, $\pi_{H}:\mathbb{S}^3\to\mathbb{S}^2$ ($\pi_H$ is the Hopf map), first developed in \cite{lagraa} (see also \cite{selene,rosa-vit}) and further related to a geometrical framework in \cite{vit-kust}. The main point underlying \cite{lagraa} is the construction of one particular deformation of $\mathbb{R}^3$, called $\mathbb{R}^3_\lambda$ in \cite{lagraa}, as a subalgebra of $\mathbb{R}^4_\theta$, the algebra of (suitable) functions of $\mathbb{R}^4\simeq\mathbb{C}^2$, says $\mathcal{F}(\mathbb{R}^4)$, endowed with the Wick-Voros product \cite{wick-voros}, a variation of the Moyal product. Hence, the star-product defining the deformation $\mathbb{R}^3_\lambda$ is essentially induced from the Wick-Voros product by the embedding of $\mathbb{R}^3_\lambda$ in $\mathbb{R}^4_\theta$ obtained from the Hopf map. It turns out that the above construction has a nice commutative geometrical analog in term of the Kustaanheimo-Stiefel map \cite{kust-stief} which can be extended to a noncommutative setting, as shown in \cite{vit-kust}. Indeed, in the commutative setting, by observing that{\footnote{The subscript $_{/0}$ means that $0$ is excluded.}} $\mathbb{R}^4_{/0}\simeq\mathbb{S}^3\times \mathbb{R}^+$ and $\mathbb{R}^3_{/0}\simeq\mathbb{S}^2\times \mathbb{R}^+$ hold true, the Hopf map $\pi_H$ can be extended to $\pi_{KS}:\mathbb{R}^4_{/0}\to\mathbb{R}^3_{/0}$, using $\mathbb{S}^3\simeq SU(2)$,  where the map $\pi_{KS}$ defines the Kustaanheimo-Stiefel fibration \cite{kust-stief}, a principal fibration with structure group $U(1)$. Then, from $\pi_{KS}$ follows the embedding of $\mathcal{F}(\mathbb{R}^3_{/0})$ in $\mathcal{F}(\mathbb{R}^4_{/0})$, while the restriction to non zero elements can be finally removed. It appears that $\pi_{KS}$ can be used to define a good notion of integral on $\mathcal{F}(\mathbb{R}^3)$ from the natural integral on $\mathcal{F}(\mathbb{R}^4)$ together with derivations on $\mathcal{F}(\mathbb{R}^3)$ from a suitable projection of those on $\mathcal{F}(\mathbb{R}^4)$. For a discussion, see \cite{valentino-marmo}. This scheme can be extended to the noncommutative setting, as shown in \cite{vit-kust}, where a natural trace on $\mathbb{R}^3_\lambda$ has been proposed as well as Laplacians with (noncommutative analog) of radial dynamics (see however \cite{galikov}). We will discuss these Laplacians at the end of the paper. The approach used in the present paper is in some sense more algebraic and does not relies explicitly on some embedding of deformations of $\mathbb{R}^3$ into some deformed $\mathbb{R}^4$. Instead, it exploits standard algebraic properties stemming from the relationship pointed out in \cite{wal-16} between those deformations of $\mathbb{R}^3$ and the convolution algebra of $SU(2)$, a particular group algebra. It is important to note that the present framework supports a natural trace simply related to the $SU(2)$ Haar measure which can be verified to be consistently related to the trace/integral on $\mathbb{R}^3_\lambda$ constructed in \cite{vit-kust}.\\
\vfill\eject

The paper is organized as follows. In Section \ref{section2}, we find that $\mathfrak{su}(2)$ Lie algebras of coordinate operators related to quantum spaces with $\mathfrak{su}(2)$ noncommutativity can be conveniently represented by $SO(3)$-covariant poly-differential  $^*$-representations. A comparison with other differential non $^*$-representations that appeared in the literature is presented. In Section \ref{subsection31}, we show that the so-called quantized plane waves obtained from the action of a suitable quantization map on the usual exponential functions can be almost determined from the combination of the usual polar decomposition of operators with an additional constraint stemming from the Wigner theorem for $SU(2)$. Subsequent standard computations detailed in Section \ref{subsection32} give rise rather easily to the explicit expression for the quantized plane waves which are found to depend on two Volterra integrals. Properties of the quantization maps are discussed in Section \ref{subsection33}. In Section \ref{subsection34}, we focus on a particular subfamily of $^*$-representations and show that the resulting star-product is equivalent to a closed star-product for the trace functional defined by the usual Lebesgue integral $\tr=\int d^3x$. This product coincides with the Kontsevich product related to the Poisson manifold dual to the finite dimensional Lie algebra $\mathfrak{su}(2)$. The Section \ref{section4} is devoted to a discussion of the results. In particular, we indicate a convenient way to extend the present construction to other semi-simple but non simply connected Lie groups, such as $SL(2,\mathbb{R})$, by making use of results from group cohomology with value in an abelian group that would replace the constraints stemming from the simple Wigner theorem used in the construction of Section \ref{subsection31}. Finally, we present noncommutative scalar field theories with quartic interactions which are free from (perturbative) UV/IR mixing as a consequence of the combination of the analysis of \cite{jpw-16} with the results of Section \ref{subsection34}.\\

\section{$\mathfrak{su}(2)$-noncommutativity and differential $^*$-representations}\label{section2}
In this section, we construct families of poly-differential representations for a coordinate operator algebra satisfying $\mathfrak{su}(2)$ Lie algebra commutation relations. Insisting on the preservation of the involutive structures of the algebra modeling the corresponding noncommutative spaces forces us to focus on $^*$-representations. We find that $SO(3)$-covariant $^*$-representations, stemming from the natural action of $SU(2)/\mathbb{Z}_2$ on the algebra modeling the quantum spaces with $\mathfrak{su}(2)$ noncommutativity, are of the form
\begin{equation} \nonumber
\hat{x}_\mu = x^\alpha \left[ f(\Delta) \delta_{\alpha \mu} + g(\Delta) \partial_\alpha \partial_\mu + i\theta \varepsilon_{\alpha \mu}^{\hspace{11pt} \rho} \partial_\rho \right] + 
\ell(\Delta) \partial_\mu ,
\end{equation}
see \eqref{general_rep} below, where $\Delta=\partial^\mu\partial_\mu$ is the usual $SO(3)$-invariant Laplacian on $\mathbb{R}^3$ and $f$, $g$ and $\ell$ are functionals of $\Delta$ which satisfy a system of differential functional equations, \eqref{1st_condition}, \eqref{2nd_condition} below. Examples are briefly discussed, some being compared to representations used recently in the literature.\\
A word of caution: To each element of the above family of $^*$-representations corresponds a star-product, hence a deformed $\mathbb{R}^3$ space, both depending on $f$, $g$, $\ell$ (and $\theta$), namely $\mathbb{R}^3_{\theta,f,g,\ell}:=(\mathcal{F}(\mathbb{R}^3),\star_{\theta,f,g,\ell})$. To simplify the notations, we will denote generically any space $\mathbb{R}^3_{\theta,f,g,\ell}$ as $\mathbb{R}^3_{\theta}$. The corresponding $^*$-representation should be clear from the context.
\subsection{The master equations}\label{subsection21}
We start from $\mathbb{A}[\hat{X}_\mu]$, the abstract operator algebra generated by the self-adjoint coordinate operators $\hat{X}_\mu$ satisfying
\begin{equation}
[\hat{X}_\mu,\hat{X}_\nu]=i2\theta\varepsilon_{\mu \nu}^{\hspace{11pt} \rho}\hat{X}_\rho,\quad  \mu,\nu,\rho=1,2,3 \label{su(2)-basic},
\end{equation}
where $\theta\in\mathbb{R}$, $\theta>0${\footnote{For the present algebraic manipulations, one can safely work at the level of the enveloping algebra of $\mathfrak{su}(2)$. See eqn. \eqref{algroupinclus}.}}. We denote by ``$^*$" the usual (canonical) involution on $\mathbb{A}[\hat{X}_\mu]$, hence a $^*$-algebra (with $\hat{X}_\mu^*=\hat{X}_\mu$), and by $\mathcal{S}(\mathbb{R}^3)$ and $\mathcal{M}(\mathbb{R}^3)$ respectively the algebra of Schwartz functions on $\mathbb{R}^3$ and its multiplier algebra. In the following, $\langle\cdot,\cdot\rangle$ denotes the usual hermitian product defined, for any $f,g\in\mathcal{M}(\mathbb{R}^3)$, by $\langle f,g \rangle:=\int d^3x\ {\bar{f}}(x) g(x)$ where ${\bar{f}}(x)$ is the complex conjugate of $f(x)$.\\
Let $\pi:\mathbb{A}[\hat{X}_\mu]\to\mathcal{L}(\mathcal{M}(\mathbb{R}^3))$, 
\begin{equation}
\pi:\hat{X}_\mu\mapsto\pi(\hat{X}_\mu)\dreq\hat{x}_\mu(x,\partial),\quad \mu=1,2,3, \label{defxhat}
\end{equation}
be a differential representation, assumed to be an algebra homomorphism, where $\mathcal{L}(\mathcal{M}(\mathbb{R}^3))$ is the set of linear operators acting on $\mathcal{M}(\mathbb{R}^3)$. Hence, \eqref{su(2)-basic} implies that
\begin{equation}
[\hat{x}_\mu,\hat{x}_\nu]=i2\theta\varepsilon_{\mu \nu}^{\hspace{11pt} \rho}\hat{x}_\rho,\label{su2hat}
\end{equation}
therefore transferring the $\mathfrak{su}(2)$ Lie algebra structure onto the relevant set of differential operators.\\
In view of further applications to NCFT, in particular to make possible the implementation of reasonable reality conditions in the construction of Lagrangians, we consider from now on the more restricted class of $^*$-representations, namely those fulfilling, for any $a\in\mathbb{A}[\hat{X}_\mu]$, 
\begin{equation}
\pi(a^*)=(\pi(a))^\dag  \ ,
\end{equation}
where the symbol ``$\dag$" denotes the usual adjoint operation on $\mathcal{L}(\mathcal{M}(\mathbb{R}^3))$. It is convenient to consider representations of the form \cite{cosmogol5}
\begin{equation}
\hat{x}_\mu=x_\nu\varphi^\nu_{\hspace{3pt}\mu}(\partial)+\chi_\mu(\partial) \ , \label{defphi}
\end{equation}
where the functionals $\varphi^\nu_{\hspace{3pt}\mu}(\partial)$ and $\chi_\mu(\partial)$ are viewed as formal expansions in the usual derivatives of $\mathbb{R}^3$, $\partial_\mu$, $\mu=1,2,3$.\\
By making use of the algebraic relation
\begin{equation}
[x_\lambda , h(x,\partial)] = - \frac{\partial h}{\partial (\partial^\lambda)},\label{commut-h}
\end{equation}
valid for any functional $h$ of $x_\mu$ and $\partial^\mu$, one finds that \eqref{defphi} obeys the $\mathfrak{su}(2)$ Lie algebra structure \eqref{su2hat} provided the following functional differential equations
\begin{align} 
\frac{\partial \varphi^\lambda_{\hspace{3pt}\mu}}{\partial (\partial^\rho)} \varphi^\rho_{\hspace{3pt}\nu} - \frac{\partial \varphi^\lambda_{\hspace{3pt}\nu}}{\partial (\partial^\rho)} \varphi^\rho_{\hspace{3pt}\mu} &= i2\theta\varepsilon_{\mu \nu}^{\hspace{11pt} \rho} \varphi^\lambda_{\hspace{3pt}\rho} \ ,\label{phidifferential}\\
\frac{\partial\chi_\mu}{\partial(\partial^\rho)}\varphi^\rho_{\hspace{3pt}\nu}-\frac{\partial\chi_\nu}{\partial(\partial^\rho)}\varphi^\rho_{\hspace{3pt}\mu} &= i2\theta\varepsilon_{\mu \nu}^{\hspace{11pt} \rho}\chi_\rho \ , \label{chidifferential}
\end{align}
hold true. \\

The above relations \eqref{phidifferential}, \eqref{chidifferential} generate infinitely many solutions for the representation $\pi$ \eqref{defxhat} defined by \eqref{defphi}. Among these, we now select the $^*$-representations which is achieved by demanding, for any $f,g\in\mathcal{M}(\mathbb{R}^3)$,
\begin{equation}
\langle f,\hat{x}_\mu g\rangle=\langle \hat{x}_\mu f,g \rangle\label{selfadjx},
\end{equation}
i.e the operators $\hat{x}_\mu$, $\mu=1,2,3$ to be self-adjoint so that $\hat{x}_\mu^\dag=\hat{x}_\mu$. Therefore, one can write, for any $f,g\in\mathcal{M}(\mathbb{R}^3)$,
\begin{align}
\langle f,\hat{x}^\dag_\mu g \rangle &= \langle(x_\alpha\varphi^\alpha_{\hspace{3pt}\mu}(\partial)+\chi_\mu(\partial))f,g \rangle=\langle f,{\bar{\varphi}}^\alpha_{\hspace{3pt}\mu}(-\partial)x_\alpha g\rangle+\langle f,{\bar{\chi}}_\mu(-\partial)g\rangle\nonumber\\
&=\langle f,x_\alpha{\bar{\varphi}}^\alpha_{\hspace{3pt}\mu}(-\partial) g\rangle+\langle f,\frac{\partial{\bar{\varphi}}^\alpha_{\hspace{3pt}\mu}(-\partial)}{\partial(\partial^\alpha)}g\rangle+\langle f,{\bar{\chi}}_\mu(-\partial)g\rangle,\label{decadix}
\end{align}
where we used 
\begin{equation}
\partial^\dag_\mu=-\partial_\mu,\ h^\dag(\partial)={\bar{h}}(-\partial)\ , \label{dagstandard}
\end{equation}
for any functional $h(\partial)$ depending on the derivatives, together with \eqref{commut-h} and the definition of the Hilbert product $\langle\cdot,\cdot\rangle$. \\
Hence, eqn. \eqref{decadix} satisfies \eqref{selfadjx} provided
\begin{align}
{\bar{\varphi}}_{\alpha\mu}(-\partial) &= \varphi_{\alpha\mu}(\partial),\label{selfadj-gene1}\\ \frac{\partial{\bar{\varphi}}_{\alpha\mu}(-\partial)}{\partial(\partial_\alpha)} &= \chi_\mu(\partial)-{\bar{\chi}}_\mu(-\partial)
\label{selfadj-gene2}.
\end{align}
From \eqref{selfadj-gene1}, one readily infers that $\varphi_{\alpha\mu}$ must have the following decomposition:
\begin{equation} \label{phi_dagger}
\varphi_{\alpha \mu} (\partial) = \Phi_{\alpha \mu}(\partial) + i \Psi_{\alpha \mu}(\partial) \ ,
\end{equation}
with the real functional $\Phi_{\alpha \mu}$ (resp. $\Psi_{\alpha \mu}$) of even (resp. odd) degree in $\partial$.\\ For further convenience, we collect below the 4 master equations \eqref{phidifferential}, \eqref{chidifferential}, \eqref{selfadj-gene1}, \eqref{selfadj-gene2} determining the poly-differential $^*$-representations
\begin{align}
i 2\theta \varphi_{\alpha \rho} &= \varepsilon_{\rho}^{\hspace{4pt} \mu \nu} \frac{\partial \varphi_{\alpha \mu}}{\partial (\partial_\beta)} \varphi_{\beta \nu},\label{master1}\\
\varphi^\dagger_{\alpha\rho} &= \varphi_{\alpha \rho} \label{master2}\\
i 2\theta \chi_\rho &= \varepsilon_{\rho}^{\hspace{4pt} \mu \nu} \frac{\partial \chi_\mu}{\partial (\partial_\alpha)} \varphi_{\alpha \nu}, \label{master3} \\
\frac{\partial \varphi^\dagger_{\alpha \rho}}{\partial (\partial_\alpha)} &= \chi_\rho - \chi_\rho^\dagger, \label{master4}
\end{align}
where we have used the algebraic relation $\delta_{\mu \gamma} \delta_\nu^{\hspace{4pt} \sigma} - \delta_\mu^{\hspace{4pt} \sigma} \delta_{\nu \gamma} = \varepsilon_{\mu \nu}^{\hspace{11pt} \rho} \varepsilon_{\rho \gamma}^{\hspace{11pt} \sigma}$ into \eqref{phidifferential} and \eqref{chidifferential} to produce \eqref{master1} and \eqref{master3}.\\
\subsection{$SO(3)$-equivariant $^*$-representations}\label{subsection22}
Owing to the fact that $\mathbb{R}^3_\theta$ supports a natural action of $SU(2)/\mathbb{Z}_2\simeq SO(3)$, since
\begin{equation}
\mathbb{R}^3_\theta\subsetneq U(\mathfrak{su}(2))\cong \mathbb{A}[\hat{X}_\mu]/[\hat{X}_\mu,\hat{X}_\nu]\label{algroupinclus}
\end{equation}
as algebras (see \cite{wal-16,jpw-16}), where $U(\mathfrak{su}(2))$ is the universal enveloping algebra of $\mathfrak{su}(2)$, we focus now on the particular class of $SO(3)$-equivariant{\footnote{This is called $SO(3)$-covariance in the Physics literature. }} representations $\pi$ which therefore correspond to $\varphi_{\alpha\mu}$ of the form
\begin{equation} 
\varphi_{\alpha \mu}(\partial) = \alpha(\Delta) \delta_{\alpha \mu} + \beta(\Delta) \left( \frac{1}{3} \delta_{\alpha \mu} \Delta - \partial_\alpha \partial_\mu \right) + \gamma(\Delta) \varepsilon_{\alpha\mu}^{\hspace{11pt} \rho} \partial_\rho,\label{phi_so3}
\end{equation}
stemming from a simple application of the Schur-Weyl decomposition theorem\footnote{See for instance in H. Weyl,  \textit{Classical groups}, Princeton University Press, 1946.} for $SO(3)$, where $\alpha$, $\beta$ and $\gamma$ are $SO(3)$-invariant functionals depending on the Laplacian $\Delta$, to be determined in a while. It will be further assumed that $\alpha$ and $\beta$ (resp. $\gamma$) are real (resp. purely imaginary) functionals so that \eqref{phi_dagger} is satisfied. In the same spirit, we write $\chi_\mu$ as
\begin{equation} 
\chi_\mu(\partial) = \ell (\Delta) \partial_\mu \ \label{polynomial_chi},
\end{equation}
where $\ell(\Delta)$ is a complex $SO(3)$-invariant functional to be determined.\\

For computational convenience we rewrite \eqref{phi_so3} as:
\begin{equation} 
\varphi_{\alpha \mu} (\partial) = f(\Delta) \delta_{\alpha \mu} + g(\Delta) \partial_\alpha \partial_\mu + i h(\Delta) \varepsilon_{\alpha\mu}^{\hspace{11pt} \rho} \partial_\rho,\label{polynomial_phi}
\end{equation}
where the {\it{real}} functionals $f(\Delta)$, $g(\Delta)$, and $h(\Delta)$ are defined by:
\begin{equation}
f(\Delta) \deq \alpha(\Delta) + \frac{\beta(\Delta)}{3} \Delta \ , g(\Delta) \deq - \beta(\Delta) \ , h(\Delta) \deq - i \gamma(\Delta).
\end{equation}
Plugging \eqref{polynomial_phi} in the first master equation \eqref{master1}, we easily obtain for the right-hand side ($RHS$) of \eqref{master1}:
\begin{equation} 
RHS=-2\theta h \varepsilon_{\alpha\mu}^{\hspace{11pt} \rho} \partial_\rho + 2i \theta \left( f \delta_{\alpha \mu} + g \partial_\alpha \partial_\mu \right),\label{RHS}
\end{equation}
and for the left-hand side ($LHS$):
\begin{align}
LHS=-(2ff' &+ 2gf' \Delta - gf + h^2) \varepsilon_{\alpha\mu}^{\hspace{11pt} \rho} \partial_\rho + \nonumber \\
&+ 2 i \left[\left(hf + gh' \Delta^2 + fh' \Delta \right) \delta_{\alpha \mu} - \left(gh' \Delta + fh' - gh \right) \partial_\alpha \partial_\mu \right], \label{LHS}
\end{align}
where $f'$ denotes the derivative of $f$ with respect to its argument. Thus, identifying \eqref{LHS} with \eqref{RHS}, the first master equation \eqref{master1} reduces to a system of three differential equations:
\begin{align}
(f+g \Delta)h' - (h - \theta)g &= 0 \ , \label{relat1} \\
(f+g \Delta)h' \Delta + (h - \theta)f &= 0 \ , \label{relat2} \\
2(f+g \Delta)f' + (h - 2\theta)h - gf &= 0 \ . \label{relat3}
\end{align}
Linear combination of the two first equations of that system leads to
\begin{equation} \label{relat6}
(f+g \Delta)(h - \theta) = 0 \ ,
\end{equation}  
highlighting the possible choices for differential representations satisfying $\mathfrak{su}(2)$ commutation relations. Namely, with either $f+g\Delta=0$ or $h=\theta$. \\
Likewise, combining \eqref{polynomial_phi} and \eqref{polynomial_chi} with \eqref{master3} and \eqref{master4}, one obtains respectively
\begin{align}
h &= \theta \label{relat4} , \\
2(f+g\Delta)' + 2g &= \ell + \ell^\dagger \label{relat5}.
\end{align}
Before giving the general form of $SO(3)$-covariant $^*$-representations $\hat{x}_\mu$ we are going to focus on next sections, one comment is in order. One can see that, as long as $\chi_\mu = 0$, \eqref{master3} gives no information for admissible solutions for $f$, $g$ and $h$. In particular, \eqref{relat4} does not necessarily hold true. Then, in the case $h\neq\theta$, one has $f+g\Delta=0$ from \eqref{relat6}, and the system \eqref{relat1}-\eqref{relat3} together with \eqref{relat5} admit two solutions\footnote{When $\chi_\mu=0$, \eqref{relat5} prevents $f+g\Delta=0$ and $h=\theta$ to be satisfied at the same time.}
\begin{align}
\hat{x}_\mu &= 0 \ , \\
\hat{x}_\mu &= i 2\theta \varepsilon_{\sigma \mu}^{\hspace{11pt} \rho} x^\sigma \partial_\rho \ ,
\end{align}
that we disregard since (introducing the left-action of operator on functions ``$\rhd$") $\hat{x}_\mu \rhd 1 \neq x_\mu$ and $\hat{x}_\mu \rhd f(x) \rightarrow 0$ when $\theta \rightarrow 0$ for arbitrary function $f$ (see section \ref{section3}). \\
Below, we consider the case $h=\theta$ (for $\chi_\mu\ne 0$ or $\chi_\mu=0$). \\

Finally, let us go back to representations satisfying the whole set of equations \eqref{relat1}-\eqref{relat5}. By observing that $h=\theta$ solves both \eqref{relat1} and \eqref{relat2} trivially, one finds the following family of $^*$-representations \eqref{defphi} (with $SO(3)$-covariance):
\begin{equation} \label{general_rep}
\hat{x}_\mu = x^\alpha \left[ f(\Delta) \delta_{\alpha \mu} + g(\Delta) \partial_\alpha \partial_\mu + i\theta \varepsilon_{\alpha \mu}^{\hspace{11pt} \rho} \partial_\rho \right] + \ell(\Delta) \partial_\mu,
\end{equation}
where the functionals $f(\Delta)$, $g(\Delta)$ and $\ell(\Delta)$ satisfy  
\begin{align}
2\left[(f+g\Delta)' + g \right] &= \ell + \ell^\dagger \ , \label{1st_condition} \\
2(f+g\Delta)f' &= gf + \theta^2 \ , \label{2nd_condition}
\end{align}
$f,g$ real.
\subsection{Discussion and example}\label{subsection23}
We now briefly discuss the above result. First, let us go back to the case of non $^*$-representations for which only the equations \eqref{relat1}-\eqref{relat3} are relevant, assuming for a while $\chi_\mu=0$. It can be easily seen that this latter system admits an interesting solution by noticing that $h=\theta$ solves both \eqref{relat1} and \eqref{relat2}, as already noticed above, while \eqref{relat3} gives rise to a Riccati equation
\begin{equation} \label{riccati}
2g' = \left(\frac{2R'(\Delta)}{\Delta} - \frac{\theta^2}{\Delta R(\Delta)} \right) - \frac{3}{\Delta}g + \frac{1}{R(\Delta)} g^2 \ ,
\end{equation}
with the constraint 
\begin{equation}
f+g\Delta\dreq R(\Delta), 
\end{equation}
where $R(\Delta)$ is some given real functional.\\
In the special case
\begin{equation}
f+g\Delta=1 \ , \label{kv-cond}
\end{equation}
one easily recovers the expression for the poly-differential (however non $^*$-)representation considered in \cite{KV-15}. Using \eqref{kv-cond}, \eqref{riccati} reduces to
\begin{equation}
2t \frac{dG}{dt} + 3\left(G(t)+1 \right) - \frac{t}{6} G^2(t) = 0 \label{equadiff-reduced} ,
\end{equation}
where we have defined 
\begin{equation}
g(\Delta)\deq\frac{\theta^2}{3} G(2\theta^2 \Delta) \ , \label{kv-def}
\end{equation}
in order to make connection with the notations and conventions of \cite{KV-15}. The solution of \eqref{equadiff-reduced} is:
\begin{equation}
G(t) = -6\sum_{n=1}^\infty \frac{2^n B_{2n}}{(2n)!} t^{n-1} \ ,\label{kv-g}
\end{equation}
where $B_n$ are Bernoulli numbers.\\

Let us now consider the case of $^*$-representations. From the above discussion, these latter must satisfy \eqref{master3}-\eqref{master4} which determine $\chi_\mu$ once $\varphi_{\mu\nu}$ fulfilling \eqref{master2} is obtained from \eqref{master1}. Hence \eqref{relat1}-\eqref{relat3} must be supplemented by \eqref{relat4}, \eqref{relat5} so that the whole system reduces to \eqref{1st_condition} and \eqref{2nd_condition}. In particular, when \eqref{kv-cond} still holds true, \eqref{1st_condition} and \eqref{2nd_condition} imply that
\begin{equation}
l+l^\dag=2g(\Delta) \ ,\label{kv-chi} 
\end{equation}
in which $g(\Delta)$ is still given by \eqref{kv-def}, \eqref{kv-g}. \\
Hence, the poly-differential representation used in \cite{KV-15} can be extended to a $^*$-representation which, by further assuming that $l(\Delta)$ is a real functional\footnote{We are going to see Subsection \ref{332} that, in view of \eqref{polar-dec}, $\ell$ is necessarily real.}, is defined by the following expression
\begin{equation}
\hat{x}_\mu=x^\alpha\left[(1-g(\Delta)\Delta)\delta_{\alpha\mu}+g(\Delta)\partial_\alpha\partial_\mu+
i\theta\varepsilon_{\alpha\mu}^{\hspace{11pt}\rho}\partial_\rho\right]+g(\Delta)\partial_\mu \ , \label{kv-star}
\end{equation}
with $g(\Delta)$ given by \eqref{kv-def}, \eqref{kv-g}.\\
In the next section, we determine explicitly the expression for the star-product corresponding to $^*$-representations defined by \eqref{general_rep}-\eqref{2nd_condition}.
\section{Quantization maps and related star-products}\label{section3}
\subsection{A natural group cohomological setting}\label{subsection31}
We start by characterizing the quantization map $Q$ and the related star-product we use in the following. We look for an invertible map which is a $^*$-algebra morphism
\begin{equation}
Q: (\mathcal{M}(\mathbb{R}^3),\star) \to (\mathcal{L}(\mathcal{M}(\mathbb{R}^3)),\cdot)\label{quant-map},
\end{equation}
where $\star$ (resp. the dot ``$\cdot$") denotes the star-product (resp. the product between poly-differential operators, omitted from now on), such that
\begin{equation}
f\star g \deq Q^{-1}\left(Q(f)Q(g)\right),\ Q(1)=\bbone,\ Q(\bar{f})=\left(Q(f)\right)^\dag \ .\label{alg-morph}
\end{equation}
Hence, one can write, for any $f,g\in\mathcal{M}(\mathbb{R}^3)$,
\begin{equation}
(f\star g)(x)=\int \frac{d^3p}{(2\pi)^3}\frac{d^3q}{(2\pi)^3}\tilde{f}(p)\tilde{g}(q)Q^{-1}\left(Q(e^{ipx})Q(e^{iqx}) \right) \ , \label{star-definition}
\end{equation}
where $\tilde{f}(p)=\int d^3xf(x)e^{-ipx}$ is the Fourier transform of $f$ (same for $\tilde{g}$) so that the star-product is fully characterized once
\begin{equation}
E_p(\hat{x}) \deq Q(e^{ipx}) \ , \label{nc-planew}
\end{equation}
together with the inverse map $Q^{-1}$ are determined. \\

For a given $^*$-representation, we observe that the determination of the inverse map $Q^{-1}$ can be conveniently carried out by enforcing the condition
\begin{equation}
Q(f)\rhd1=f(x) \ ,\label{Q-unitaction}
\end{equation}
for any $f\in\mathcal{M}(\mathbb{R}^3)$, where the symbol ``$\rhd$" denotes the left action of operators, hence 
\begin{equation}
Q^{-1}\left(Q(f)\right)=Q(f) \rhd 1.
\end{equation}
In the following, the construction of the star-product is performed by requiring that \eqref{Q-unitaction} holds true.\\

Next we recall, as shown in \cite{wal-16}, that the noncommutative algebra $\mathbb{R}^3_\theta$ generated by \eqref{su2hat} is the $SU(2)$ Fourier transform of (hence isomorphic to) the convolution algebra of $SU(2)$, $(L^2(SU(2)), \bullet)$, where the symbol $\bullet$ denotes the associative convolution product on $SU(2)$ given, for any functions $f,g\in L^1(SU(2))$, by
\begin{equation}
(f\bullet g)(u)=\int_{SU(2)} d\mu(t)f(t) g(t^{-1}u),
\end{equation}
in which $d\mu(t)$ is the $SU(2)$ Haar measure. Hence, it is natural to view $E_p(\hat{x})$, eqn. \eqref{nc-planew}, as stemming from a map $E:SU(2)\to\mathcal{L}(\mathcal{M}(\mathbb{R}^3))$, with
\begin{equation}
E:g\mapsto E(g):=E_p(\hat{x}), \label{themap}
\end{equation}
and
\begin{equation}
E(g^\dag)=E^\dag(g) \ , \label{themapdag}
\end{equation}
for any $g\in SU(2)$.\\
In the following, the quantity $E(g)\equiv E_p(\hat{x})$ is called "quantized plane waves". \\
Now, from the general polar decomposition of an operator, one can write
\begin{equation}
E(g)=U(g)|E(g)| \ , \label{polar-dec}
\end{equation}
where $U:SU(2)\to\mathcal{L}(\mathcal{M}(\mathbb{R}^3))$ is a unitary operator and $|E(g)|:=\sqrt{E^\dag(g)E(g)} \neq 0$ denotes as usual the absolute value of $E(g)$. In view of the Stone's theorem, it is legitimate to parametrize the unitary operator involved in \eqref{polar-dec} as 
\begin{equation}
U(g)=e^{i\xi_g^\mu\hat{x}_\mu}, 
\end{equation}
where $\xi_g^\mu\in\mathbb{R}$, to be determined in a while in terms of the functionals $f,\ g,\ \ell$ entering eqn. \eqref{general_rep}. Hence, $U(g)$ can be viewed as an element of $SU(2)$ and the Baker-Campbell-Hausdorff formula for $\mathfrak{su}(2)$ applies between exponential functions $e^{i\xi^\mu_g\hat{x}_\mu}$, namely
\begin{equation}
e^{i\xi_{g_1}\hat{x}}e^{i\xi_{g_2}\hat{x}}=e^{iB(\xi_{g_1},\xi_{g_2})\hat{x}} \ ,
\end{equation}
where $B(\xi_{g_1},\xi_{g_2})$ is an infinite expansion satisfying
\begin{equation}
B(\xi_{g_1},\xi_{g_2})=-B(-\xi_{g_2},-\xi_{g_1}) , \ B(\xi_g,0)=\xi_g \ .
\end{equation}
Since $U(g)$ and $E(g)$ define representations of $SU(2)$, one has for any $g_1,g_2\in SU(2)$
\begin{equation}
U(g_1)U(g_2)=U(g_1g_2) \ , \label{projectif-su2}
\end{equation}
which holds up to {\it{unitary equivalence}} as a mere application of the Wigner theorem to $SU(2)$, while we demand
\begin{equation}
E(g_1)E(g_2)=\Omega(g_1,g_2)E(g_1g_2)\label{projectif-caracter}
\end{equation}
where $\Omega(g_1,g_2)$ will be determined in a while. In particular, \eqref{projectif-caracter} leads to
\begin{equation}
E(g^\dag)E(g)=\Omega(g^\dag,g)\bbone
\end{equation}
for any $g\in SU(2)$, where we used $E(g^\dag g)=E(\bbone)=\bbone$. Therefore,
\begin{equation}
\vert E(g)\vert=\sqrt{\Omega(g^\dag,g)}\bbone.
\end{equation}
Hence, one has 
\begin{equation}
\omega_g:=\sqrt{\Omega(g^\dag,g)} \in \mathbb{R},\ \omega_g>0,
\end{equation}
together with
\begin{equation}
[|E(g)|,U(g)]=0. \label{cond-central}
\end{equation}
Combining \eqref{cond-central} with \eqref{polar-dec}, one easily obtains:
\begin{equation}
E(g_1)E(g_2)=|E(g_1)||E(g_2)|U(g_1g_2)=|E(g_1)||E(g_2)||E(g_1g_2)|^{-1}E(g_1g_2) \ , \label{intermediaire}
\end{equation}
where the second equality stems from \eqref{projectif-su2}.\\
Using the expression for $|E(g)|$, one can rewrite \eqref{intermediaire} as
\begin{equation}
E(g_1)E(g_2)=(\omega_{g_1}\omega_{g_2}\omega^{-1}_{g_1g_2})E(g_1g_2) \ , \label{planew-multiplic}
\end{equation}
where
\begin{equation}
E(g_1g_2)=\omega_{g_1g_2}e^{iB(\xi_{g_1},\xi_{g_2})\hat{x}}.
\end{equation}
\\
At this point, some comments are in order:\\

First, we note that the general form for $E(g)$ derived above may be guessed by performing a brute force computation of $e^{ik\hat{x}}\rhd 1$. This is given in the Appendix \ref{zassen-comput}.\\

Next, it can be easily realized that \eqref{planew-multiplic} insures automatically the associativity of the $SU(2)$ group product and therefore the associativity of the related star-product \eqref{star-definition}. This comes from the fact that $\Omega(g_1,g_2):=\omega_{g_1}\omega_{g_2}\omega^{-1}_{g_1g_2}$ obeys a 2-cocycle relation, namely 
\begin{equation}
\Omega(g_1,g_2)\Omega(g_1g_2,g_3)=\Omega(g_1,g_2g_3) \Omega(g_2,g_3),
\end{equation}
for any $g_1,g_2,g_3\in SU(2)$.\\

Finally, from eqn. \eqref{projectif-su2}, it follows that any unitary equivalent representations, says $U$ and $U^\prime$, give rise to unitary equivalent products. Indeed, by unitary equivalence, one can write 
\begin{equation}
U^\prime(g)=e^{i\gamma(g)}U(g)=e^{i\gamma(g)}e^{i\xi_g\hat{x}} 
\end{equation}
where $\gamma$ is a real function. Then, it can be easily verified that one can write the following relation
\begin{equation}
T(f\star^\prime g)=Tf\star Tg,
\end{equation}
which defines obviously an equivalence relation between the two star-products, where the map $T$ is defined from
\begin{equation}
E^\prime_k(\hat{x})\equiv Q^\prime(e^{ikx}):=Q\circ T(e^{ikx})=e^{i\gamma(k)}Q(e^{ikx})=e^{i\gamma(k)}E_k(\hat{x}).
\end{equation}
This can be rephrased by stating that the star-product which will be determined in a while is essentially unique up to unitary equivalence. \\
A similar conclusion applies for any other semi-simple and simply connected Lie group $G$, assuming of course that a suitable $^*$-representation for its Lie algebra similar to \eqref{su2hat} has been determined. This reflects the fact that $H^2(G,\mathbb{R}/\mathbb{Z})$, the second cohomology group of $G$ with value in $U(1)\simeq\mathbb{R}/\mathbb{Z}$ is trivial, implying that any unitary map $U:G\to End(\mathcal{H})$ (where $\mathcal{H}$ is some suitable Hilbert (representation) space) satisfies 
\begin{equation}
U(g_1)U(g_2)=\Gamma(g_1,g_2)U(g_1g_2)\label{projective-rep}
\end{equation}
for any $g_1,g_2\in G$ where the map $\Gamma:G\times G\to U(1)$ is simply a coboundary (i.e a trivial cocycle) which therefore takes the form 
\begin{equation}
\Gamma(g_1,g_2)=e^{i\gamma(g_1)}e^{i\gamma(g_2)}e^{-i\gamma(g_1g_2)}. 
\end{equation}
Hence, as in the case of $G=SU(2)$, one can set $\Gamma(g_1,g_2)=\bbone$, any other star-product obtained from $\Gamma(g_1,g_2)\ne\bbone$ being unitary equivalent to the one corresponding to $\Gamma(g_1,g_2)=\bbone$.\\
Notice that eqn. \eqref{projective-rep} defines a projective representation of $G$. Recall that (inequivalent) projective representations of any (connected) Lie group $G$ are classified by the 2-cocycles pertaining to $H^2(G,U(1))$. Basic properties of the group cohomology with values in a $G$-module that are relevant here are recalled in the Appendix \ref{cohomologydisc} for the sake of completeness.\\

It appears that semi-simple but non simply connected groups may have non trivial $H^2(G,U(1))$, leading to the appearance of different classes of star-products that we briefly discuss Section \ref{section4}.
\subsection{Determination of the quantized plane waves}\label{subsection32}
We are now in position to characterize from the family of $^*$-representations $\hat{x}_\mu$ defined by \eqref{general_rep}-\eqref{2nd_condition}, the explicit expression for the quantized plane waves, \eqref{themap},
\begin{equation}
E_p(\hat{x})=\omega(p)e^{i\xi(p)\hat{x}} \label{generalform-ncexpo} ,
\end{equation}
discussed Subsection \ref{subsection31}, by actually computing $\omega(p)$ and $\xi(p)$ whose general expressions are given by two Volterra integrals, \eqref{volterra-omega} and \eqref{volterra-xi} respectively.
\subsubsection{Computation of $\xi(p)$}\label{section321}
Let's first derive the expression for $\xi$. From the definition of ${E}_p(\hat{x})$, one has:
\begin{equation} \label{expo_appendix}
e^{i\xi(p)\hat{x}}\rhd 1 = \frac{e^{ipx}}{\omega(p)} \ .
\end{equation}
Then, the action of $e^{-i\xi(p) \hat{x}} \partial_\mu e^{i\xi(p) \hat{x}}$ on 1 gives:
\begin{align*}
e^{-i\xi(p) \hat{x}} \partial_\mu e^{i\xi(p) \hat{x}} \rhd 1 &= e^{-i\xi(p) \hat{x}} \partial_\mu \rhd \frac{e^{ipx}}{\omega(p)} \\
&= e^{-i\xi(p) \hat{x}} \rhd (ip_\mu) \frac{e^{ipx}}{\omega(p)} \\
&= (ip_\mu)  e^{-i\xi(p) \hat{x}} e^{i\xi(p) \hat{x}} \rhd 1 \ .
\end{align*}
Since $e^{-i\xi(p) \hat{x}} e^{i\xi(p) \hat{x}} \equiv \bbone$, one gets the following (operator) identity:
\begin{equation} \label{operator_identity}
e^{-i\xi(p) \hat{x}} \partial_\mu e^{i\xi(p) \hat{x}} = (ip_\mu) \bbone \ ,
\end{equation}
which is satisfied for any 3-momentum $p\in \mathbb{R}^3$, hence holds true by rescaling $p \mapsto \lambda p$, with $\lambda \in \mathbb{R}$, namely 
\begin{equation}
e^{-i\xi(\lambda p) \hat{x}} \partial_\mu e^{i\xi(\lambda p) \hat{x}} = (i\lambda p_\mu) \bbone.
\end{equation}
Then, making use of the derivative w.r.t. $\lambda$ on this last expression gives $(ip_\mu) \bbone$ for the RHS and, for the LHS:
\begin{equation}
\frac{d}{d\lambda} \left[ e^{-i\xi(\lambda p) \hat{x}} \partial_\mu e^{i\xi(\lambda p) \hat{x}} \right] = i \frac{d}{d\lambda} \left[ \xi^\nu(\lambda p) \right] \left( e^{-i\xi(\lambda p) \hat{x}} \varphi_{\mu \nu}(\partial) e^{i\xi(\lambda p) \hat{x}} \right) \ ,
\end{equation}
where we have used the identity $[\partial_\mu, \hat{x}_\nu] = [\partial_\mu,x^a\varphi_{a\nu}] = \varphi_{\mu \nu}$. Since $\varphi_{\mu\nu}$ is a function of $\partial$ only, it is straightforward to show (in view of \eqref{operator_identity}) that:
\begin{equation}
e^{-i\xi(\lambda p) \hat{x}} \varphi_{\mu \nu}(\partial) e^{i\xi(\lambda p) \hat{x}} = \varphi_{\mu\nu}(i\lambda p) \bbone \ , 
\end{equation}
then (by identification of LHS and RHS):
\begin{equation}\label{diff-xi}
\varphi_{\mu\nu}(i\lambda p) \frac{d}{d\lambda} \left[ \xi^\nu(\lambda p) \right] = p_\mu ,
\end{equation}
which permits us to determine $\xi(p)$ by solving a first order differential equation. \\

To do so, we first have to invert $\varphi_{\mu\nu}$. From our discussion Section \ref{section2}, we search solution of the form
\begin{equation} \label{phi-inverse-general}
(\varphi^{-1})_{\mu\nu}(\partial) \deq X(\Delta) \delta_{\mu \nu} + Y(\Delta) \partial_\mu \partial_\nu + Z(\Delta) \varepsilon_{\mu \nu}^{\hspace{11pt} \rho} \partial_\rho \ ,
\end{equation}
such that $\varphi_{\mu\nu} (\varphi^{-1})^{\nu\sigma} = \delta_\mu^{\hspace{3pt}\sigma}$. A standard computation leads to the following system:
\begin{align}
fX - i\theta \Delta Z &= 1 \ , \nonumber \\
(f+\Delta g)Y + gX + i\theta Z &= 0 \ , \\ 
fZ + i\theta X &= 0 \ , \nonumber
\end{align}
which admits the following unique solution (assuming $f^2\neq\theta^2\Delta$)\footnote{In the case $f^2-\theta^2\Delta=0$, $\varphi_{\mu\nu}$ is not invertible.}:
\begin{align}
X(\Delta) &= \frac{f(\Delta)}{f^2(\Delta)-\theta^2\Delta}  \ , \label{X-coord} \\
Y(\Delta) &= - \frac{2f'(\Delta)}{f^2(\Delta)-\theta^2\Delta}\ , \label{Y-coord} \\
Z(\Delta) &= - \frac{i \theta}{f^2(\Delta)-\theta^2\Delta} \ , \label{Z-coord}
\end{align}
where we have used equation \eqref{2nd_condition} to simplify the expression for $Y$. Then:
\begin{equation}
(\varphi^{-1})^{\mu\nu}(ip) = \frac{1}{f^2+\theta^2 p^2} \left( f \delta^{\mu\nu} + 2f'p^\mu p^\nu + \theta \varepsilon^{\mu \nu \rho} p_\rho \right) \ , \label{phi-inverse}
\end{equation}
where $f$ and its derivative, defined by \eqref{general_rep}-\eqref{2nd_condition}, are functions of $(-p^2)$.\\

Finally, the expression for $\xi(p)$ follows directly by integrating $d\xi^\mu = (\varphi^{-1})^{\mu\nu}_{\vert_{i\lambda p}} p_\nu d\lambda$ between 0 and 1. Namely, one has
\begin{equation} \label{solution-xi}
\xi^\mu(p) = \int_0^1 d\lambda (\varphi^{-1})^{\mu\nu}_{\vert_{i\lambda p}} p_\nu \ ,
\end{equation}
where $\varphi^{-1}_{\vert_y}$ refers to the function $\varphi^{-1}$ evaluated on $y$, whose expression is given by \eqref{phi-inverse}, and where we have used $\xi_\mu(0)=0$ stemming from $E(\bbone)=E_0(\hat{x})=\bbone$. \\
At this stage, one remark is in order. One can easily verify that, by setting 
\begin{equation}
f=1+p^2g \ ,
\end{equation}
which corresponds to the case \eqref{kv-cond}, \eqref{kv-star} discussed Section \ref{section2}, one has
\begin{equation}
(\varphi^{-1})^{\mu\nu}p_\nu = p^\mu,
\end{equation}
thus recovering the expected result $\xi^\mu(p) = p^\mu$ (cf. eqn. \eqref{expo-vitale}). \\

One can rewrite \eqref{solution-xi} as a Volterra integral by using \eqref{phi-inverse-general} and the change of variable $t=-\lambda^2 p^2$:
\begin{equation} \label{volterra-xi}
\xi^\mu(p) = \int_{-p^2}^0 \frac{dt}{2p \sqrt{-t}} \ \left[X(t) + t Y(t)\right] p^\mu \ ,
\end{equation} 
Note that $\xi_\mu$ is an injective antisymmetric real-valued function. 
\subsubsection{Computation of $\omega(p)$}\label{332}

Now, it remains to determine $\omega(p)$ in order to fully characterize $E_p(\hat{x})$. Let us rescale $p\mapsto \lambda p$, $\lambda \in \mathbb{R}$ in \eqref{expo_appendix}. On one hand, one has:
\begin{equation}
\frac{d}{d \lambda} \left[ e^{i\xi(\lambda p) \hat{x}} \right] = i \frac{d}{d\lambda} \left[ \xi^\mu(\lambda p) \right] \hat{x}_\mu e^{i\xi(\lambda p) \hat{x}} = i (\varphi^{-1})^{\mu\nu}_{\vert_{i\lambda p}}  p_\nu \hat{x}_\mu e^{i\xi(\lambda p) \hat{x}} \ ,
\end{equation}
where we used \eqref{solution-xi} for the second equality. Then,
\begin{align}
\frac{d}{d \lambda} \left[ e^{i\xi(\lambda p) \hat{x}} \right] \rhd 1 &= i (\varphi^{-1})^{\mu\nu}_{\vert_{i\lambda p}}  p_\nu \left(x^\alpha \varphi_{\alpha \mu}(\partial) + \chi_\mu(\partial) \right) \rhd \frac{e^{i\lambda px}}{\omega(\lambda p)} \nonumber \\
&= i (\varphi^{-1})^{\mu\nu}_{\vert_{i\lambda p}}  p_\nu \left(x^\alpha \varphi_{\alpha \mu}(i\lambda p) + \chi_\mu(i\lambda p) \right) \frac{e^{i\lambda px}}{\omega(\lambda p)} \nonumber \\
&= i \left( x^\nu + \chi_\mu (\varphi^{-1})^{\mu\nu}_{\vert_{i\lambda p}} \right) p_\nu \frac{e^{i\lambda px}}{\omega(\lambda p)} \ .
\end{align}
On the other hand, one can write
\begin{equation}
\frac{d}{d \lambda} \left[ \frac{e^{i\lambda px}}{\omega(\lambda p)} \right] = \left( ix^\nu p_\nu - \frac{1}{\omega(\lambda p)} \frac{d}{d \lambda} \left[ \omega(\lambda p) \right] \right) \frac{e^{i\lambda px}}{\omega(\lambda p)}.
\end{equation}
Now, from the action of the derivation operation
\begin{equation}
\frac{dg}{d\lambda} \deq \lim_{\epsilon \rightarrow 0} \frac{g(\lambda + \epsilon)-g(\lambda)}{\epsilon},
\end{equation}
on $g(\lambda,x)=\hat{A}(\lambda)f(x)$ for some suitable operator $\hat{A}$ and function $f$, one obtains 
\begin{equation}
\frac{dg}{d\lambda} = \lim_{\epsilon \rightarrow 0} \left( \frac{\hat{A}(\lambda + \epsilon)-\hat{A}(\lambda)}{\epsilon} \right) f(x)\ . 
\end{equation}
Hence, one can write
\begin{equation}
\frac{d}{d\lambda} \left[ \hat{A} f(x) \right] = \frac{d\hat{A}}{d\lambda} f(x).
\end{equation}
From this, we easily obtain
\begin{equation}
\frac{d}{d\lambda} \left[ e^{i\xi(\lambda p) \hat{x}} \rhd 1 \right] = \frac{d}{d\lambda} \left[ e^{i\xi(\lambda p) \hat{x}} \right] \rhd 1 \ , 
\end{equation}
and
\begin{equation}
i \left( x^\nu + \chi_\mu (\varphi^{-1})^{\mu\nu}_{\vert_{i\lambda p}} \right) p_\nu = ix^\nu p_\nu - \frac{1}{\omega(\lambda p)} \frac{d}{d \lambda} \left[ \omega(\lambda p) \right] \ ,
\end{equation}
which gives rise to the following differential equation:
\begin{equation}
\frac{1}{\omega(\lambda p)} \frac{d}{d \lambda} \left[ \omega(\lambda p) \right] = - i \chi_\mu (\varphi^{-1})^{\mu\nu}_{\vert_{i\lambda p}} p_\nu \ ,
\end{equation}
admitting the following solution
\begin{equation} \label{solution-omega}
\omega(p) = e^{-i \int_0^1 d\lambda \ \chi_\mu(i\lambda p) (\varphi^{-1})^{\mu\nu}_{\vert_{i\lambda p}} p_\nu} \ .
\end{equation}

This latter equation \eqref{solution-omega} can also be written as a Volterra equation using \eqref{phi-inverse-general} and performing the change of variable $t=-\lambda^2 p^2$:
\begin{equation} \label{volterra-omega}
\omega(p) = e^{\int_{-p^2}^0 dt \left[X(t) + t Y(t)\right]\ell(t)} \ .
\end{equation}
Note that $\omega$ is symmetric function, namely $\omega(-p)=\omega(p)$. \\

At this point, one important comment is in order. According to the discussion Subsection \ref{subsection31}, $\omega(p)$ must be a positive real quantity. By observing that $X$ and $Y$, \eqref{X-coord}, \eqref{Y-coord}, are real, it follows that $\ell$ has to be a real functional, $\ell^\dag=\ell$.\\
Hence, eqn. \eqref{1st_condition} entering the definition of the family of $^*$-representations \eqref{general_rep} reduces to
\begin{equation}
\ell=(f+g\Delta)^\prime+g,
\end{equation}
which therefore constraints the expression for $\ell$ once $f$ and $g$ satisfying \eqref{2nd_condition} are determined.

\subsection{Quantization maps and $^*$-representations}\label{subsection33}
According to the discussion Subsection \ref{subsection31}, the quantization map $Q$, \eqref{quant-map}, as well as the star-product, \eqref{star-definition}, are determined once the map $E$, \eqref{themap}, is known. In practice, this is done by first choosing a $^*$-representation fulfilling \eqref{general_rep}-\eqref{2nd_condition}, then computing $\omega$ and $\xi$ using \eqref{volterra-xi} and \eqref{volterra-omega}. In this section, we show that $Q$ \textit{(whose inverse is given by \eqref{Q-unitaction})} cannot be the Weyl quantization map $W$ corresponding to the symmetric ordering of operators whenever the poly-differential representation $\hat{x}_\mu$ is a $^*$-representation as defined by \eqref{defphi}. In other words, from eqn. \eqref{weyl-expon}, that means we cannot find $^*$-representation such that $\omega(p)=1$, $\xi_\mu(p) = p_\mu$, for all $p \in \mathbb{R}^3$, and $W(e^{ipx})\rhd 1 = e^{ipx}$.\\

To see that, first recall that the Weyl map is defined for any function $f$ with corresponding expansion 
\begin{equation}
f(x)=\sum \limits_{n}\alpha_{\mu_1\mu_2...\mu_n}x^{\mu_1}x^{\mu_2}...x^{\mu_n} 
\end{equation}
and for any representation $\hat{x}$, by the map $W$:
\begin{align}
W(f)&\deq \sum_n\alpha_{\mu_1\mu_2...\mu_n}\langle\hat{x}^{\mu_1}\hat{x}^{\mu_2}...\hat{x}^{\mu_n}\rangle_S \ ,\label{weylmap}\\
\langle \hat{x}^{\mu_1}\hat{x}^{\mu_2}...\hat{x}^{\mu_n}\rangle_S &\deq \sum_{P_n\in\mathfrak{S}_n}\frac{1}{n!}P_n(\hat{x}^{\mu_1}\hat{x}^{\mu_2}...\hat{x}^{\mu_n}) \ , \label{symm-prod}
\end{align}
where the sum in the symmetric product \eqref{symm-prod} runs over all the permutations of $n$ operators $\hat{x}_\mu$. Hence, one easily realizes that
\begin{align}
W(e^{ipx})&= e^{ip\hat{x}}\label{weyl-expon},\\
W(f)&= \int\frac{d^3p}{(2\pi)^3}\tilde{f}(p)e^{ip\hat{x}}\label{weyl-mapfinal}.
\end{align}
Now, when the operators $\hat{x}_\mu$ are defined by \eqref{general_rep}-\eqref{2nd_condition}, a simple computation yields
\begin{align}
\langle \hat{x}_{\mu_1} \hat{x}_{\mu_2} \rangle_S \rhd 1 &= x_{\mu_1} x_{\mu_2} + \ell_0 \delta_{\mu_1 \mu_2} \ , \label{decadix1} \\
\langle \hat{x}_{\mu_1} \hat{x}_{\mu_2} \hat{x}_{\mu_3} \rangle_S \rhd 1 &= x_{\mu_1} x_{\mu_2} x_{\mu_3} + \frac{2}{3} \left(f_1 + g_0 + \frac{3}{2} \ell_0 \right) x_{(\mu_1} \delta_{\mu_2 \mu_3)} \ , \label{decadix2} \\
\langle \hat{x}_{\mu_1} \hat{x}_{\mu_2} \hat{x}_{\mu_3} \hat{x}_{\mu_4} \rangle_S \rhd 1 &= x_{\mu_1} x_{\mu_2} x_{\mu_3} x_{\mu_4} + \frac{2}{3} \left( f_1 \ell_0 +  g_0 \ell_0 + \frac{3}{2} \ell_0^2 + 3 \ell_1 \right) x_{(\mu_1} \delta_{\mu_2 \mu_3)} + \nonumber\\
& +\frac{4}{3} \left( f_1 + g_0 + \frac{15}{16} \ell_0 \right) \left( \delta_{\mu_1 (\mu_2} x_{\mu_3} x_{\mu_4)} + x_{\mu_1} x_{(\mu_2} \delta_{\mu_3 \mu_4)} \right) \ , \dots \label{First_terms},
\end{align}
in which the $f_i$'s (resp. $g_i$, $\ell_i$) are coefficients appearing in the expansion of $f(\Delta)$ (resp. $g(\Delta)$, $\ell(\Delta)$). Then, one concludes that unless the functionals $f(\Delta)$, $g(\Delta)$, $l(\Delta)$ satisfy some relation which implies the vanishing of all the terms involving Kronecker symbol ($\sim\delta_{\mu\nu}$) in the RHS of eqns. \eqref{decadix1}-\eqref{First_terms}, an arbitrary $\hat{x}_\mu$ in the family of $^*$-representations \eqref{general_rep}-\eqref{2nd_condition} is such that
\begin{equation}
W(e^{ipx})\rhd1\ne e^{ipx},\label{Q-nonunit}
\end{equation}
and thus $W(f)\rhd1\ne f(x)$ so that the condition \eqref{Q-unitaction} is not satisfied. Note that, at least for the family of solutions we will consider, eqn. \eqref{cond-central}, such a relation implying the above vanishing does not exist as we now show.\\

Let us assume that there exists a $^*$-representation for which $W(e^{ipx}) \rhd 1 = e^{ipx}$. From the discussion Subsection \ref{subsection31}, one then would have $W(e^{ipx})=\omega(p)e^{i\xi(p)\hat{x}}$ with $\omega(p)=1$ and $\xi^\mu(p) = p^\mu$. But this would imply from \eqref{solution-xi} and \eqref{solution-omega} 
\begin{equation}
\int_{-p^2}^0\ dt(X(t)+tY(t))\ell(t)=0,\label{integralone}
\end{equation}
\begin{equation}
\int_{-p^2}^0\ \frac{dt}{2t\sqrt{-t}}(X(t)+tY(t))=1\label{integraltwo}.
\end{equation}
Whenever $\chi(t)\ne0$, it can be easily checked that \eqref{integralone} and \eqref{integraltwo} cannot be simultaneously satisfied. Indeed, \eqref{integralone} implies $(X(t)+tY(t))\ell(t)=0$ and therefore
\begin{equation}
X(t)+tY(t)=0 \ ,
\end{equation}
which clearly contradicts \eqref{integraltwo}.\\
When $\chi(t)=0$, \eqref{integraltwo} implies the following Riccati equation
\begin{equation}
2tf^\prime=t\theta^2+f-f^2 \label{riccati-lol},
\end{equation}
whose solution does not solve \eqref{1st_condition}, \eqref{2nd_condition} (for $\ell=0$).
Hence, we conclude that there is no poly-differential $^*$-representation \eqref{general_rep} that could fit with the Weyl quantization map. \\

As a remark, we notice that the fact that $^*$-representations \eqref{general_rep} with $\chi=0$ cannot give rise to quantized plane waves of the general form \eqref{generalform-ncexpo} fulfilling \eqref{weyl-expon}, is already apparent in e.g eqn. \eqref{diff-xi}. Indeed, recall that $\xi_\mu(p)=p_\mu$ holds true whenever $f+gt=1$, as discussed in Subsection \ref{subsection32}, which however is not compatible with \eqref{1st_condition}-\eqref{2nd_condition}.\\ 
One concludes that  the only poly-differential representation compatible with the Weyl quantization map within the framework of Subsection \ref{subsection31} is the non $^*$-representation discussed in Subsection \ref{subsection23} whose form is given by
\begin{equation}
\hat{x}_\mu=x^\alpha\left((1-g\Delta)\delta_{\alpha\mu}+g\partial_\alpha\partial_\mu+i\theta\varepsilon_{\alpha\mu\rho}\partial_\rho\right)
\end{equation}
where $g$ is given by \eqref{kv-def}, \eqref{kv-g}.\\

\subsection{Closed star-product}\label{subsection34}

From now on, we focus on the example discussed in Subsection \ref{subsection23}, for which the differential $^*$-representation \eqref{defxhat} is given by \eqref{kv-star} $$\hat{x}_\mu=x^\alpha\left[(1-g(\Delta)\Delta)\delta_{\alpha\mu}+g(\Delta) \partial_\alpha\partial_\mu + i\theta\varepsilon_{\alpha\mu}^{\hspace{11pt}\rho}\partial_\rho\right]+g(\Delta)\partial_\mu \ ,$$ where $g(\Delta) = - \sum_{n=1}^\infty \frac{(2\theta)^{2n} B_{2n}}{(2n)!} \Delta^{n-1}$, which (formally) converges to:
\begin{equation} \label{closed_g}
g(\Delta) = - \Delta^{-1} \left( \theta\sqrt{\Delta} \coth(\theta\sqrt{\Delta}) - 1 \right) .
\end{equation}
Now, recall that $g$ is obtained from a Riccati equation given by $2tf'=t\theta^2+f-f^2$ where $f=1-tg$ (see \eqref{riccati}, \eqref{kv-cond}). Then, it follows from \eqref{X-coord} and \eqref{Y-coord} that $X(t)+t Y(t)=1$. Hence, using \eqref{volterra-omega} and \eqref{volterra-xi}, one finds that the corresponding quantized plane waves take the form
\begin{equation}
E_p(\hat{x}) = \omega(p) e^{ip\hat{x}} \ ,
\end{equation}
with 
\begin{equation}
\omega(p) = \exp \left[ \int_{-p^2}^0 g(y) dy\right], 
\end{equation}
and $g$ still given by \eqref{closed_g}. \\
Passing from hyperbolic to trigonometric functions and performing the change of variable $x=\theta \sqrt{y}$, one can rewrite $\omega$ as
\begin{equation}
\omega(p) = \exp\left[2 \int^{\theta |p|}_0 \left( \cot(x) - \frac{1}{x} \right) dx \right].
\end{equation}
Upon integrating this latter expression, one finally obtains the following quantized plane waves
\begin{equation} \label{pw-example}
Q(e^{ipx})\equiv E_p(\hat{x}) = \left( \frac{\sin(\theta |p|)}{\theta |p|} \right)^2 e^{ip\hat{x}}.
\end{equation}
According to the discussion of Subsection \ref{subsection31}, the corresponding $\star$-product, denoted by $\star_Q$, is readily obtained from
\begin{equation}
e^{ipx} \star_Q e^{iqx} = \mathcal{W}^2(p,q) e^{iB(p,q)x} \ ,
\end{equation}
with
\begin{equation}
\mathcal{W}(p,q) \deq \frac{|B(p,q)|}{\theta |p||q|}\frac{\sin(\theta |p|)\sin(\theta |q|)}{\sin(\theta |B(p,q)|)} \ ,\label{theweight}
\end{equation}
and $B(p,q)$, stems from the Baker-Campbell-Haussdorff formula for $\mathfrak{su}(2)$, as introduced in Subsection \ref{subsection31}. One therefore has
\begin{equation}
(f\star_Q g)(x) = \int \frac{d^3p}{(2\pi)^3}\frac{d^3q}{(2\pi)^3}\tilde{f}(p)\tilde{g}(q) \mathcal{W}^2(p,q) e^{iB(p,q)x} \ ,
\end{equation}
for any $f,g \in \mathcal{M}(\mathbb{R}^3)$.\\

Now, define a new quantization map $\mathcal{K}:\mathcal{M}(\mathbb{R}^3) \to \mathcal{L}(\mathcal{M}(\mathbb{R}^3))$ as
\begin{equation}
\mathcal{K} \deq Q \circ H,
\end{equation}
where the operator $H$ acting on the functions of the algebra $\mathcal{M}(\mathbb{R}^3)$ is given by
\begin{equation} \label{Kontsevich}
H \deq \frac{\theta \sqrt{\Delta}}{\sinh(\theta \sqrt{\Delta})},
\end{equation}
and such that
\begin{equation}
H(f\star_\mathcal{K}g)=H(f)\star_QH(g),
\end{equation}
for any $f,g \in \mathcal{M}(\mathbb{R}^3)$, which defines obviously an equivalence relation between the star-products $\star_Q$ and $\star_\mathcal{K}$.\\
A standard calculation yields
\begin{equation}
\mathcal{K}(e^{ipx}) = \frac{\sin(\theta |p|)}{\theta |p|} e^{ip\hat{x}} \label{checkpoint2}.
\end{equation}
Hence the corresponding $\star$-product $\star_\mathcal{K}$, which is ($H$-)equivalent to $\star_Q$, can be obtained from 
\begin{equation}
e^{ipx} \star_\mathcal{K} e^{iqx} = \mathcal{W}(p,q) e^{iB(p,q)x},
\end{equation}
and one can write, for any $f,g \in \mathcal{M}(\mathbb{R}^3)$,
\begin{equation}
(f\star_\mathcal{K}g)(x)=\int \frac{d^3p}{(2\pi)^3}\frac{d^3q}{(2\pi)^3}\tilde{f}(p)\tilde{g}(q) \mathcal{W}(p,q) e^{iB(p,q)x} \ , \label{kontsev-product}
\end{equation}
where $\mathcal{W}(p,q)$ is still given by \eqref{theweight}.\\

This star-product $\star_\mathcal{K}$, \eqref {kontsev-product}, coincides with the Kontsevich product \cite{deform}. This has been derived for $\mathbb{R}^3_\theta$ within a different approach in \cite{q-grav} (see also \cite{KV-15}), namely
\begin{equation}
\mathcal{K}=W\circ j^{\frac{1}{2}}(\Delta) \ , \label{deriv-konts}
\end{equation}
where $W$ is the Weyl quantization map and 
\begin{equation}
j^{\frac{1}{2}}(\Delta)=\frac{\sinh(\theta \sqrt{\Delta})}{\theta \sqrt{\Delta}} \ , \label{duflo}
\end{equation}
is the Harish-Chandra map \cite{harish}, \cite{duflo}. Recall that $\star_\mathcal{K}$ is closed for the trace functional defined by the Lebesgue integral on $\mathbb{R}^3$, namely
\begin{equation}
\int d^3 x (f \star_\mathcal{K} g)(x) = \int d^3 x f(x) g(x).
\end{equation}
Comparing \eqref{Kontsevich} and \eqref{duflo}, one infers
\begin{equation}
j^{\frac{1}{2}}(\Delta)=H^{-1},
\end{equation}
hence $H$ \eqref{Kontsevich} is the inverse of the Harish-Chandra map. Notice that, by using \eqref{pw-example} combined with \eqref{checkpoint2}, one has 
\begin{equation}
\mathcal{K}(e^{ipx})\rhd\bbone=\frac{\theta|p|}{\sin(\theta|p|)}e^{ipx},
\end{equation}
while \eqref{checkpoint2} and \eqref{deriv-konts} yield
\begin{equation}
W(e^{ipx})=\frac{\theta|p|}{\sin(\theta|p|)}\mathcal{K}(e^{ipx})=e^{ip\hat{x}}.
\end{equation}
\section{Discussion}\label{section4}
Let us summarize and discuss the results of this paper. As we have shown, abstract $\mathfrak{su}(2)$ Lie algebras of coordinates underlying popular quantum spaces with $\mathfrak{su}(2)$ noncommutativity can be conveniently represented by poly-differential involutive representations hence preserving the involutions of the various algebraic structures modeling these quantum spaces. Their natural $SO(3)$-covariance singles out a particular family of involutive representations defined by eqns. \eqref{general_rep}-\eqref{2nd_condition}. A brief comparison with non-involutive representations used in the literature is done.\\

Given an involutive representation, the related star-product is obtained once the action of a suitable quantization map on the usual exponential functions (plane waves) is defined, giving rise to the so-called quantized plane waves. We show that their expression is mostly constrained by the general polar decomposition of operators combined with additional constraints stemming from the Wigner theorem for $SU(2)$. This leads to the general expression \eqref{generalform-ncexpo} for the quantized plane waves which is shown to depend on two functions of the momenta. A standard computation then leads to the explicit expressions for these functions in term of two Volterra integrals. \\

We note, by the way, that the star-product used in e.g \cite{lagraa} to define a particular deformation of $\mathbb{R}^3$ (called $\mathbb{R}^3_\lambda$) does not belong to the general family of star-products related to \eqref{general_rep} as it can be easily verified by computing $x_\mu\star x_\nu=\hat{x}_\mu\hat{x}_\nu\triangleright 1$ using \eqref{general_rep} and comparing the result to e.g the formula (2.19) of \cite{lagraa}. Thus, the deformation defining $\mathbb{R}^3_\lambda$ does not pertain to the type of deformations we obtained in this paper. Notice that the former deformation is related to the Wick-Voros product \cite{wick-voros, galuccio, fedele} which stems from a twist. We do not know at the present time whether or not our family of star-products also admits a representation in terms of a twist. Assuming such a twist exists and has an expansion in terms of the deformation parameter, preliminary (tedious) computation of the first terms suggests that the answer is likely positive. The corresponding complete determination, which is an interesting question, is beyond the scope of this paper.\\

As shown in Section \ref{subsection31}, uniqueness up to unitary equivalence of the related star-product simply reflects the trivial structure of the projective representations for $SU(2)$. This can be translated more abstractly as a consequence of the triviality of $H^2(SU(2),U(1))$, the second cohomology group of $SU(2)$ valued in $U(1)$. Recall that group cohomology with value in an abelian group is the proper tool to investigate the so-called central extensions of a group, which is a convenient way to extend the present work to more complicated class of Lie groups. Some basic details are given in the Appendix \ref{cohomologydisc}. \\
Let us discuss this result on a more general footing. Let $\mathcal{D}$, $G$ and $\mathfrak{g}$ denote respectively a discrete group of $\mathbb{R}$ (i.e $\mathcal{D}=p\mathbb{Z}$), a Lie group and its Lie algebra. Consider the case of central extensions of $G$ by a 1-dimensional abelian group{\footnote{As a remark, notice that the 1-dimensional abelian group serving to extend $G$ is not related to any subgroup of $G$.}} $\mathcal{A}=\mathbb{R}/\mathcal{D}$. When $G$ is {\it{simply connected}}, a general result in mathematics states that 
\begin{equation}
H^2(G,\mathbb{R}/\mathcal{D})\simeq H^2_{alg}(\mathfrak{g},\mathbb{R}), 
\end{equation}
see e.g \cite{hoch-shap}, where $H^2_{alg}(\mathfrak{g},\mathbb{R})$ is the 2nd group of real cohomology of the Lie algebra $\mathfrak{g}$, which then both classify the inequivalent central extensions of $G$ by the compact group $\mathbb{R}/\mathcal{D}$. Now, when $G$ is in addition semi simple, so that $\mathfrak{g}$ is also semi simple, it is known that $H^2_{alg}(\mathfrak{g},\mathbb{R})=\{0\}$. Hence, $H^2(G,\mathbb{R}/\mathcal{D})$ is trivial implying uniqueness of the central extension of any semi-simple and simply connected Lie group by $\mathbb{R}/\mathcal{D}$. \\
When $\mathcal{D}=\mathbb{Z}$, one has $\mathbb{R}/\mathcal{D}=U(1)$. Then, the triviality of $H^2(SU(2),U(1))$ extends to any semi-simple and simply connected Lie group $G$ and any extension of the present scheme to a coordinate algebra obeying the commutation relations of the Lie algebra of $G$ can be expected to give rise to a unique (up to unitary equivalence) star-product.\\

Before discussing possible extensions of the present work to non simply connected groups, one remark is in order. Let us briefly compare our scheme to the construction based on the Hopf map mentioned in the introduction. Recall that this latter map permits one to define $\mathbb{S}^3$ (isomorphic to the Lie group $SU(2)$) as a fiber bundle over the base manifold (but not a Lie group) $\mathbb{S}^2$ with fiber being the $\mathbb{S}^1$ isomorphic to the compact Lie group $U(1)$. The structure underlying our construction is different in that the Lie group $SU(2)$ is now the base space of a (principal) fiber bundle with fiber $U(1)$, the central extension of $SU(2)$ by $U(1)$. More generally, the general structure of our construction requires the use of Lie groups (and their associated Lie algebras, these latter being related to the coordinate algebras defining the noncommutative spaces). In particular, the central extension of $G$ by $\mathbb{R}/\mathcal{D}$, says $\tilde{G}$, defines (up to technical requirements) a (principal) fiber bundle $\tilde{G}\to G$ with structure group $\mathbb{R}/\mathcal{D}$. Extensions of the Hopf scheme \cite{lagraa} using generalized Hopf fibrations{\footnote{Fibrations between spheres are only possible for $\mathbb{S}^1$, $\mathbb{S}^3$, $\mathbb{S}^7$, $\mathbb{S}^{15}$.}} (e.g $\mathbb{S}^n\to\mathbb{R}P^n$ or $\mathbb{S}^{2n+1}\to\mathbb{C}P^n$, $n>3$), if possible at all, appears to be related to a different fiber bundle structure, e.g, base spaces are not (always) Lie groups, $\mathbb{S}^n$, $n>3$ is not a Lie group.\\

In view of future generalizations \cite{PW}, we point out that semi-simple but non simply connected groups may have non trivial $H^2(G,U(1))$, leading to the appearance of inequivalent classes of star-products. Indeed, one can show that the inequivalent central extensions of a semi-simple Lie group $G$ are classified, up to some additional technical requirements, by $H^1_{\check{C}}(G,\mathbb{R}/\mathcal{D})$, where $H^\bullet_{\check{C}}$ refers to the \v{C}ech cohomology. But a standard result in algebraic topology states that
\begin{equation}
H^1_{\check{C}}(G,\mathbb{R}/\mathcal{D})\simeq Hom(\pi_1(G)\to\mathbb{R}/\mathcal{D})
\end{equation}
where the RHS denotes the group of homomorphisms from the first homology group of $G$ into $\mathbb{R}/\mathcal{D}$. Hence, the inequivalent central extensions of semi-simple but non simply connected groups $G$ by $\mathbb{R}/\mathcal{D}$ are classified by the group $Hom(\pi_1(G)\to\mathbb{R}/\mathcal{D})$.\\
Now, pick $G=SL(2,\mathbb{R})$ and $\mathcal{D}=\mathbb{Z}$ so that once more time $\mathbb{R}/\mathcal{D}$ is the compact U(1) group. From Iwazawa decomposition, one infers $SL(2,\mathbb{R})\simeq \mathbb{R}^2\times \mathbb{S}^1$ as topological spaces{\footnote{Note that the compact subgroup $\mathbb{S}^1\simeq U(1)$ in $SL(2,\mathbb{R}$) is obviously not related to the abelian group $\mathbb{R}/\mathcal{D}\vert_{\mathcal{D}=\mathbb{Z}}=U(1)$}}. Using $\pi_1(X\times Y)=\pi_1(X)\times\pi_1(Y)$ for any (topological) spaces $X$ and $Y$, one obtains
$\pi_1(SL(2,\mathbb{R}))\simeq\mathbb{Z}$ and thus $Hom(\mathbb{Z}\to U(1))\simeq U(1)$, which classifies the inequivalent extensions of $SL(2,\mathbb{R})$ by $U(1)$. \\
Notice that a simpler example is provided by $SO(3)$ for which one has $\pi_1(SO(3))=\mathbb{Z}/2\mathbb{Z}$ so that the relevant group is $Hom(\mathbb{Z}/2\mathbb{Z}\to U(1))$ and one recovers the 2 inequivalent projective representations of $SO(3)$, i.e indexed by $\Gamma=\pm\bbone$ in \eqref{projective-rep}. \\ 

Finally, focusing on a particular subfamily of $^*$-representations indexed by a single real functional of $\Delta$, the laplacian on $\mathbb{R}^3$, we have shown in Section \ref{subsection34} that the corresponding star-product $\star_\mathcal{K}$ is equivalent to the Kontsevich product related to the Poisson manifold dual to the finite dimensional Lie algebra $\mathfrak{su}(2)$, hence closed for the trace functional defined by the usual Lebesgue integral $\tr=\int d^3x$. Then, the analysis of \cite{jpw-16} can be straightforwardly adapted to scalar noncommutative field theories with functional actions (in obvious notations) given by 
\begin{eqnarray}
S_1&=&\int d^3x\big[\frac{1}{2}\partial_\mu\phi\star_\mathcal{K}\partial_\mu\phi+\frac{1}{2}m^2\phi\star_\mathcal{K}\phi+\frac{\lambda}{4!}\phi\star_\mathcal{K}\phi\star_\mathcal{K}\phi\star_\mathcal{K}\phi\big]\label{real-clasaction},\\
S_2&=&\int d^3x\big[\partial_\mu\Phi^\dag\star_\mathcal{K}\partial_\mu\Phi+m^2\Phi^\dag\star_\mathcal{K}\Phi+
{\lambda}\Phi^\dag\star_\mathcal{K}\Phi\star_\mathcal{K}\Phi^\dag\star_\mathcal{K}\Phi\big]\label{complx-clasaction},
\end{eqnarray}
which obviously admit standard (i.e commutative) massive real or complex scalar field theories with quartic interaction as formal commutative limits. From \cite{jpw-16}, one concludes in particular that \eqref{real-clasaction} and \eqref{complx-clasaction} do not have (perturbative) UV/IR mixing. Indeed, a standard analysis as in e.g \cite{jpw-16} shows that the quadratic part of the effective action for \eqref{real-clasaction} (i.e 2-point function part) receives 2 types of one-loop contributions whose typical form is given by
\begin{eqnarray}
\Gamma_2^{(I)}&=&\int d^3x\ \phi(x)\phi(x)\omega_I\label{cosmogol-alp},\\
\Gamma_2^{(II)}&=&\int \frac{d^3k_1}{(2\pi)^3}\frac{d^3k_1}{(2\pi)^3}\ \tilde{\phi}(k_1)\tilde{\phi}(k_2)\omega_{II}(k_1,k_2),
\end{eqnarray}
where $\tilde{\phi}$ is the Fourier transform of $\phi$, in which
\begin{eqnarray}
\omega_{I}&\sim&\frac{4}{\theta^2}\int\frac{d^3p}{(2\pi)^3}\ \frac{\sin^2(\frac{\theta}{2}|p|)}{p^2(p^2+m^2)}=\frac{1-e^{-\theta m}}{2m\pi\theta^2}\\
\omega_{II}&\sim&\int d^3x\frac{d^3p}{(2\pi)^3}\ \frac{1}{p^2+m^2}(e^{ipx}\star_{\mathcal{K}}e^{ik_1x}\star_{\mathcal{K}}e^{-ipx}\star_{\mathcal{K}}e^{ik_2x}),
\end{eqnarray}
up to unessential overall factors. For $\theta\ne0$, one can check that $\omega_I$ is finite (even for $m=0$) while one obtains $\omega_{II}(0,k_2)\sim\delta(k_2)\omega_I$ (with similar expression for $\omega_{II}(k_1,0)$). Besides, UV one-loop finiteness of $\omega_{II}$ can be verified in the same way as done in \cite{jpw-16}. Similar conclusions hold true for the complex scalar field case \eqref{complx-clasaction}. Then, absence of IR singularity signals the absence of perturbative UV/IR mixing. 

Let us comment this result. We first note that the absence of UV/IR mixing (together with a mild and even all order finite UV behaviour) also occurs in other NCFT built on deformations of $\mathbb{R}^3$ as well as in related gauge theory versions as shown in \cite{vitwal}, for which a factorization property of the corresponding partition functions played a salient role. This factorization property stems from the Peter-Weyl decomposition of the noncommutative algebra combined with the fact that the kinetic operators considered in these theories have a blockwise diagonal representation induced by the Peter-Weyl decomposition. Indeed, the operator algebra linked with these deformations with $\mathfrak{su}(2)$ noncommutativity is isomorphic to the convolution algebra of $SU(2)$, the isomorphism being simply defined by the $SU(2)$ Fourier transform, see \cite{wal-16}. Hence, the noncommutative space splits into an (infinite) orthogonal sum of finite noncommutative geometries, each one modeled by $\mathbb{M}_{2j+1}(\mathbb{C})$, $j\in\frac{\mathbb{N}}{2}$, and the NCFT split into an infinite tower of (matrix) field theories on finite geometries, each one having a natural cut-off provided by, says the radius $\sim j$ of the relevant ``fuzzy sphere" $\mathbb{M}_{2j+1}(\mathbb{C})$, thanks to the blockwise diagonal decomposition of the considered kinetic operators. \\
For the NCFT \eqref{real-clasaction}, \eqref{complx-clasaction} considered in this paper, the Peter-Weyl decomposition still holds true for the algebra, so that one still has a natural cut-off similar to the one mentionned above. However, investigating the perturbative properties to any order (i.e the UV behavior to any order) is more complicated than for the above mentioned NCFT, while however no IR singularity shows up as shown above. In fact, while the interaction part of the action can still be easily represented as a matrix model interaction using the canonical Peter-Weyl basis as done in \cite{vitwal}, the kinetic part involving the usual Laplacian on $\mathbb{R}^3$, despite its familiar form, would lead to a complicated expression (presumably not block-diagonal) when represented as a matrix model kinetic part, which would make the analysis of the UV behavior (at any order) more involved. Checking that the UV finitess of the present NCFT extends to any order (that we conjecture to be true) would require to use a (convenient) parametric representation of the general amplitudes combined with suitable estimates. This is beyond the scope of the paper.\\

It would be interesting to extend these NCFT to the case of noncommutative gauge theories built from differential calculi which {\it{do not}} belong to the category of usual derivation-based differential calculi \cite{mdv-jcw}. These latter are known to give rise to gauge theory models whose commutative limit do not coincide with standard gauge theories on $\mathbb{R}^3$, reflecting the fact that no (analog of) radial dependence can be accommodated in this framework. Hence, natural noncommutative Laplacians do not reduce to $\Delta$ at the commutative limit. Interesting candidates to consider would be the bicovariant differential calculus which is a natural case to consider on $\mathbb{R}^3_\theta$, see e.g \cite{magi-magi}. Such type of noncommutative differential calculus may well lead to Laplacian having the expected commutative limit. Notice that another interesting proposal has been analyzed in \cite{fedele}, which however amounts to enlarge the initial algebra by incorporating the deformation parameter itself. This permits one to define an additional 4th (and quite natural) derivation (hence a 4th ``engineering" dimension) related to the dilation. Doing this, a radial dependence is introduced in the derivations. Such a noncommutative differential calculus would presumably gives rise to some ``Laplacian" having the right commutative limit. The related gauge theories from the viewpoint of their perturbative quantum behavior would be worth investigating. Notice that in \cite{lagraa} appeared another early attempt to construct NCFT on $\mathbb{R}\times\mathbb{R}^3_\theta$ in which it is tempting to interpret the extra factor $\mathbb{R}$ as related to a ``time" direction. Then, provided a suitable radial dependence is re-installed through a suitable choice for the differential calculus (which should concern the $\mathbb{R}^3_\theta$ part of the algebra), it would then be possible to study NCFT (having a right commutative limit) on this noncommutative space(-time).

\vskip 2 true cm

{\bf{Acknowledgments:}} J.-C. Wallet warmly thanks F. Latr\'emoli\`ere for various exchanges and discussions on algebraic and group cohomological aspects connected to the present work. T. Poulain thanks COST Action MP1405 QSPACE for partial financial support. General discussions with M. Dubois-Violette are gratefully acknowledged. This work is partially supported by H2020 Twinning project No. 692194, “RBI-T-WINNING” and we also acknowledge the support by Croatian Science Foundation under the project IP-2014-09-9582.

\appendix 
\section{Computation of $e^{ip\hat{x}}\rhd\bbone$}\label{zassen-comput}
In order to have more insight on the possible expression for $E_p(\hat{x})$, \eqref{nc-planew}, \eqref{themap}, it is instructive to study the action of $e^{ip\hat{x}}$ on $1$. This can be achieved by using the Zassenhaus formula stemming from the Baker-Campbell-Hausdorff formula. Namely, for any operators $X$ and $Y$, one can write
\begin{equation}
e^{X+Y} = e^X e^Y \prod \limits_{n=2}^{\infty} e^{C_n(X,Y)} \ ,\label{zassenhaus}
\end{equation}
where $C_n(X,Y)$ is a homogeneous Lie polynomial of degree $n$ depending on $X$ and $Y$ of the form
\begin{equation}
[V_1,[\dots,[V_{n-1},V_n]\dots]] \ , V_i \in \lbrace X,Y \rbrace \ , \ i \in \lbrace 1,2,\dots n \rbrace \ .
\end{equation}
A mere application of \eqref{zassenhaus} to $e^{ip\hat{x}}$ where $\hat{x}_\mu$ is of the general form \eqref{defphi}, setting $X=q^\mu x^\alpha\varphi_{\alpha\mu}(\partial)$ and $Y=q^\mu\chi_\mu(\partial)$, gives
\begin{equation}
e^{iq^\mu(x^a \varphi_{\alpha \mu}+\chi_\mu)} = e^{iq^\mu x^a \varphi_{\alpha \mu}(\partial)} e^{iq^\mu \chi_\mu(\partial)} F_q(\partial) \label{expo-zassen}
\end{equation}
where $F_q \deq \prod \limits_{n \geq 2} e^{C_n(X,Y)}$ depends only on $\partial$ (and $q$), as it can be easily shown by induction. \\
Indeed, this is apparent from the expressions for $C_2$ and $C_3$ respectively given by 
\begin{equation}
C_2=-\frac{1}{2}[X,Y] \ , \ C_3=\frac{1}{3}[C_2,X+2Y] \ ,
\end{equation}
with $X$ and $Y$ just given above and further using relation \eqref{commut-h}. Now, assume that $C_{n}$, $n>3$, depends only on the $\partial_\mu$'s. Then, the Lie homogeneous polynomials of degree $n+1$ involves terms of the generic form $[q^\mu(x^\alpha\varphi_{\alpha\mu}(\partial)),\tilde{C}_{n}(\partial)]$ or $[q^\mu\chi_\mu(\partial),\tilde{C}_{n}(\partial)]$ where the symbol $\tilde{C}_n$ denotes generically terms involved in $C_n$ which therefore depends only on $\partial$ by assumption. But from a simple computation, one immediately obtains
\begin{equation}
[q^\mu\chi_\mu(\partial),\tilde{C}_{n}(\partial)]=0\label{2ndcommut0},
\end{equation}
while the first commutator reduces to 
\begin{equation}
[q^\mu x^\alpha\varphi_{\alpha\mu}(\partial),\tilde{C}_{n}(\partial)]=-q^\mu\frac{\partial \tilde{C}_n}{\partial(\partial_\alpha)}\varphi_{\alpha\mu},\label{1stcommut}
\end{equation}
where we used \eqref{commut-h}. It follows that $F_q$ depends only on the $\partial_\mu$'s (and $q$) as announced above.\\

We are now in position to guess a reasonable Ansatz for $E_p(\hat{x})$. From \eqref{expo-zassen}, we get:
\begin{equation} \label{expo-2}
e^{iq\hat{x}} \rhd 1 = F_q(0) e^{iq^\mu \chi_\mu(0)} \left( e^{iq^\mu x^\alpha \varphi_{\alpha \mu}(\partial)} \rhd 1 \right)= F_q(0) \left( e^{iq^\mu x^\alpha \varphi_{\alpha \mu}(\partial)} \rhd 1 \right) \ , 
\end{equation}
with $\chi_\mu(0) = 0$ from \eqref{polynomial_chi} and $F_q(0) \neq 1$ depending on $\varphi$, $\chi$, their derivatives (all evaluated in 0) and $q$. In the last equation \eqref{expo-2}, the term $e^{i q^\mu x^a \varphi_{a \mu}(\partial)} \rhd 1$ corresponds to the case mainly studied in the literature (where poly-differential representation are used), namely for representation preserving the Lie algebra structure but not the involution. \\
We see that for $^*$-representations, the addition of $\chi_\mu$ in the definition of the poly-differential representation $\hat{x}_\mu$, modifies the action of $e^{iq\hat{x}}$ on 1 by a factor depending (in a non trivial way) on $\varphi_{a \mu}$, $\chi_\mu$ and their derivatives. However, $e^{i q^\mu x^a \varphi_{a \mu}(\partial)} \rhd 1$ will still give (in general\footnote{As discussed in Subsection \ref{subsection33}, there is actually only one representation for which $e^{ip\hat{x}} \rhd 1 = e^{ipx}$. However, it is important to notice that the exponent $q^\mu x^a \varphi_{a \mu}(\partial)$, involved in $e^{i q^\mu x^a \varphi_{a \mu}(\partial)} \rhd 1$, is not self-adjoint as shown Subsection \ref{subsection21}. Thus, $e^{i q^\mu x^a \varphi_{a \mu}(\partial)}\notin SU(2)$. This is in particular the case for the representation studied in \cite{KV-15}.}) a result different from the commutative plane wave $e^{iqx}$. \\
However, in the case $\hat{x}_\mu$ given by \eqref{kv-star}, $e^{i q^\mu x^a \varphi_{a \mu}(\partial)} \rhd 1 = e^{ipx}$ and \eqref{expo-2} reduces to
\begin{equation} \label{expo-vitale}
e^{iq\hat{x}} \rhd 1 = F_q(0) e^{ipx} \ ,
\end{equation}
which leads to the following quantized plane waves
\begin{equation} \label{pw-vitale}
E_q(\hat{x}) = \frac{e^{iq\hat{x}}}{F_q(0)} \ .
\end{equation}

In view of \eqref{pw-vitale}, it is tempting to extend its validity to the general family of representations \eqref{general_rep}-\eqref{2nd_condition}. Therefore, we look for quantized plane waves \eqref{nc-planew} of the form
\begin{equation}
E_p(\hat{x})=\omega(p)e^{i\xi(p)\hat{x}}\label{ansatz-pw},
\end{equation}
for any $^*$-representation of the form \eqref{general_rep}-\eqref{2nd_condition}, which satisfies
\begin{equation}
E_p(\hat{x})\rhd1=e^{ipx}\label{invers-pw},
\end{equation}
so that $Q^{-1}(Q(f))=Q(f)\rhd1=f$ for any $f\in\mathcal{M}(\mathbb{R}^3)$.\\
We further require that
\begin{equation}
\bar{\xi}(p)=\xi(p)\label{real-phase} \ ,
\end{equation}
what implies that the operators of the form $e^{i\xi(p)\hat{x}}$ are unitary, therefore belong to $SU(2)$ and thus insures that the operator product between the $E_p(\hat{x})$ is associative. This implies that the related star-product is associative, as a mere consequence of the factorisation of the prefactor in \eqref{ansatz-pw} combined with the BCH formula for $\mathfrak{su}(2)$.
\section{Basics on group cohomology}\label{cohomologydisc}
Let $G$ be a (connected) Lie group. It turns out that the $U(1)$-valued 2-cocycles mentioned in the Subsection \ref{subsection31} can be actually interpreted as 2-cocycles relation for a differential on some cochains groups with values in a $G$-module $\mathcal{A}$, i.e an abelian Lie group with an action $\rho$ of $G$ on $\mathcal{A}$,
\begin{equation}
\rho:G\times\mathcal{A}\to\mathcal{A}, 
\end{equation}
which will be assumed to be trivial in the following, namely $\rho(g,a)=a$ for any $a\in\mathcal{A}$, $g\in G$. This pertains to the framework of the cohomology $H^\bullet(G,\mathcal{A})$, i.e the cohomology of a (connected) group $G$ with value in $\mathcal{A}$ that we now briefly describe. For more mathematical details together with applications in (commutative) quantum field theory, see e.g \cite{brown}, \cite{stor-wal}. \\

It is the cohomology of the complex $(\mathcal{C}^\bullet(G,\mathcal{A}),\delta)$ where the graded space 
\begin{equation}
\mathcal{C}^\bullet(G,\mathcal{A})=\bigoplus_{p\in\mathbb{N}}\mathcal{C}^p(G,\mathcal{A})
\end{equation}
is built from the $p$-cochain groups 
\begin{equation}
\mathcal{C}^p(G,\mathcal{A})=\{\Omega:\underbrace{G\times G...\times G}_p\to\mathcal{A} \}
\end{equation}
for any $p\in\mathbb{N}$, with coboundary operator 
\begin{equation}
\delta:\mathcal{C}^p(G,\mathcal{A})\to\mathcal{C}^{p+1}(G,\mathcal{A}) 
\end{equation}
defined for any $\Omega\in\mathcal{C}^p(G,\mathcal{A})$ and any $g_1,g_2,...,g_{p+1}\in G$ by
\begin{eqnarray}
(\delta\Omega)(g_1,...,g_{p+1})&=&g_1\Omega(g_2,...,g_{p+1})+\sum_{k=1}^p(-1)^k\Omega
(g_1,...,g_kg_{k+1},...,g_{p+1})\nonumber\\
&+&(-1)^{p+1}\Omega(g_1,...,g_{p})\label{cobordism}.
\end{eqnarray}
One can check that $\delta$ is nilpotent, i.e 
\begin{equation}
\delta^2=0. 
\end{equation}
Denoting by $\mathcal{Z}^p(G,\mathcal{A})$ and $\mathcal{B}^p(G,\mathcal{A})$ respectively the groups of $p$-cocycles ($\delta\Omega=0$) and $p$-coboundaries ($\Omega=\delta\rho$), one has 
\begin{equation}
H^p(G,\mathcal{A})=\mathcal{Z}^p(G,\mathcal{A})/\mathcal{B}^p(G,\mathcal{A}), 
\end{equation}
for any $p\in\mathbb{N}$, and 
\begin{equation}
H^\bullet(G,\mathcal{A})=\bigoplus_{p\in\mathbb{N}} H^p(G,\mathcal{A})
\end{equation}
(with in particular $H^0(G,\mathcal{A})=\mathcal{A}^G$, the set of $G$-invariant elements in $\mathcal{A}$).\\

Of particular interest here is $H^2(G,\mathcal{A})$. This latter classifies all the inequivalent central extensions of $G$ by $\mathcal{A}$. Note that group central extensions can be related to Lie algebra central extensions. It is more convenient here to use the framework of group extensions. Recall that $\mathcal{E}$ is a central extension of $G$ by $\mathcal{A}$ if $\mathcal{A}$ is (isomorphic to) a subgroup of the center of $G$ and one has the group isomorphism $G\simeq \mathcal{E}/\mathcal{A}$. Two extensions, says $\mathcal{E}_1$ and $\mathcal{E}_2$, are equivalent if one can find a group isomorphism $\psi:\mathcal{E}_1\to\mathcal{E}_2$ such that 
\begin{equation}
\pi_2\circ\psi=\pi_1 
\end{equation}
and
\begin{equation}
\psi(g_1\alpha)=\psi(g_1)\alpha
\end{equation}
for any $g_1\in\mathcal{E}_1$, $\alpha\in\mathcal{A}$, where $\pi_i:\mathcal{E}_i\to G$, $i=1,2$ denotes the canonical projection in the sense of group homomorphisms. More abstractly, $H^2(G,\mathcal{A})$ classifies all the inequivalent central extensions of $G$ by $\mathcal{A}$, encoded in the exact short sequences
\begin{equation}
\bbone\to\mathcal{A}\to\mathcal{E}\overset{\pi}{\to} G\to \bbone, 
\end{equation}
supplemented by the condition $Im(\mathcal{A})\subset\mathcal{Z}(\mathcal{E})$, where $\mathcal{Z}(\mathcal{E})$ is the center of $\mathcal{E}$. \\

Many results are known in mathematics whenever $\mathcal{A}$ is a 1-dimensional abelian group, $\mathcal{A}=\mathbb{R}/\mathcal{D}$ where $\mathcal{D}$ is a discrete subgroup of $\mathbb{R}$, which is the situation considered below. Note that in this case the central extension $\mathcal{E}$ can be actually interpreted as a principal fiber bundle over the group $G$ with structure group $\mathcal{A}=\mathbb{R}/\mathcal{D}$. We do not exploit this viewpoint in this paper. \\

When $G$ is semi simple and simply connected (which is the case of $SU(2)$), it turns out that the only central extension of $G$ by $\mathbb{R}/\mathcal{D}$ is the trivial one $G\times\mathbb{R}/\mathcal{D}$. In particular, whenever $\mathcal{D}=\mathbb{Z}$, one easily recovers the Wigner theorem from \eqref{projective-rep}, \eqref{projectif-su2} since from the above triviality one can set $\Gamma=1$ in \eqref{projective-rep} which thus defines the unique (up to unitary equivalence) unitary representation of $G$.\\

When the Lie group $G$ is semi-simple but not simply connected, the situation becomes non trivial (see the discussion in Section 4). Note the existence of relationships between central extensions of groups and central extensions of corresponding Lie algebras which however is not so well adapted to the present construction.

\vfill\eject

\end{document}